# THE NEWTONIAN MECHANICS OF DEMAND

**Max B. Mendel**[*]
Delft Center of Systems and Control
Department of Mechanical Engineering
Delft University of Technology
Delft, The Netherlands

October 26, 2023

## ABSTRACT

Economic engineering is a new field wherein economic systems are modelled in the same manner as traditional mechanical and electrical engineering systems. In this paper, we use Newton's theory of motion as the basis for the theory of demand; thereby establishing a theoretical foundation for economic engineering. We follow Newton's original development, as set forth in the Principia, to determine economic analogs to his three laws of motion. The pivotal result is an operational definition for an economic force, i.e. a want or a desire, in terms of a price adjustment. With this, we model the price effects of scarcity and trade friction in analogy with the models for the spring and damping force. In turn, we define economic benefits and surplus as analogous to the definitions of mechanical work and energy. These are then used to interpret the various types of economic equilibrium considered by economists from a mechanical perspective. The effectiveness of the analogy is illustrated by applying it to modelling the price and inventory dynamics of various economic agents —including consumers, dealers, holders, spot and futures traders— using linear time-invariant systems theory.

## 1 Introduction

Newton's laws of motion in mechanics and the law of demand in economics have several commonalities. Both are considered to be fundamental to their respective fields. Also, like Newtonian mechanics, the theory of demand is based on causation, unlike econometric approaches that are based on correlation. In mechanics, it is a change in momentum that causes a change in velocity, and in the theory of demand it is a price change that causes a change in the quantity demanded.

However, while Newton's theory forms the basis for the dynamics of a mechanical system, the theory of demand is focused on the equilibrium of an economic system. In Newtonian mechanics, the rate of change of momentum is defined to be a force and this definition determines the differential equations that describe motion through a balance of forces. An equivalent operational definition of a force is lacking in economics in general, and in the theory of demand in particular. Instead, economists justify in a metaphorical manner how presumed "forces of demand and supply" or an "invisible hand" achieve price equilibrium "in the long run" (see classics such as [1], [2], [3], and [4]).

Our purpose with this paper is to fill that gap by developing a fully dynamic theory of demand based on the format of Newton's formulation of his laws of motion. We are inspired in this by Maxwell, who developed electrodynamics in this manner. In engineering, this method of analogs has been extended to other domains including, amongst others, fluids and thermodynamics. It has also proven to be useful for modelling mixed systems and to design controllers in a manner independent of the physical domain. Further extension would leverage engineering design and automatic-control techniques to economic systems or even mixed econo-electromechanical systems.

---
[*]The author wishes to thank the students in the economic engineering group, and Coen Hutters in particular, without whose innumerable contributions this manuscript could not have been created.

In subsequent sections 2 and 3, we follow Newton's initial "Definitions" and "Laws" sections of his Principia. In Section 2, we present the economic analogs to Newton's definitions of inertia (demand), velocity (quantity demanded), momentum (price), and force (want). In Section 3, we use these definitions to formulate the economic analogs to Newton's three laws. The critical step is the analog to the second law. We interpret an economic force as a want or desire and consider it analogous to Newton's motive force. Furthermore, we interpret the force of demand (or supply) as the rate of change of the price, analogous to Newton's definition of the inertial force as the rate of change of momentum. The analog to Newton's second law equates the two, defining a want as a rate of change in price. We recognize this as the law of demand: the more someone wants something, the higher the price they are willing to pay.

Other attempts to use the method of analogs in economics consider price itself, rather than rate of change, to be analogous to a force (see [5] and the references therein). However, without a well-defined cause for price dynamics, the theory that emerges is more Aristotelian than Newtonian in nature. In particular, a Newtonian force is zero in equilibrium. If we were to accept that price is a force, then price is zero in equilibrium and this cannot be considered an accurate description of reality.

To be useful for building economic models, such a Newtonian law of demand requires the economic analogs to force laws or to constitutive relations. We refer to these as price drivers. We show in Section 4 how Hooke's law or the spring law can be considered analogous to the law of scarcity or storage and how the friction laws from mechanics are equally applicable to economics. In Section 5 we show how the theory allows one to consider the economic surplus as the analog to kinetic energy and how the various price drivers allow us to introduce other types of benefits that emerge to be analogous to other forms of energy and work. Section 6 rounds off the theoretical development with a comparison between the various concepts of economic equilibrium to those in mechanics.

A Newtonian theory of demand allows us to model and control economic systems using the same dynamical systems theory that engineers use to model electro-mechanical systems. We refer to this field as economic engineering. In Section 7, we show how even first- and second-order dynamical systems can model sophisticated economic dynamics such as price rigidities, trade cycles, and hyperbolic discounting. It should be noted that the dynamical systems approach is distinct from Forrester's system dynamics approach (see [6]), despite the similarity in names. Although stocks are established by accumulating a flow of goods, system dynamics lacks the concept of a force that may accumulate to establish a price. From a dynamical-systems perspective, such models are kinematic and static, rather than truly dynamic.

Although our primary audience are mechanical engineers, we have attempted to make the paper accessible to anyone with at least a solid high school physics background. The first two sections follow Newton's own development (in the [7] translation), which is eminently readable even today. Feynman [8] gives a modern perspective that is true to Newton's development. In Appendix A, we include the electrodynamic analog familiar to electrical engineers and a hydrodynamic analog which is readily visualized in general.

Our intent here is to fit economics to the mold of mechanics, rather than the other way around. For each mechanical concept or principle, we search for what appears to be the most appropriate economic analog and consider that the "economic-engineering" analog, even though this might be incompatible with some thinking in economics. We intentionally avoid references to any recent or specialized research in the economic literature and, instead, focus on making a connection with the classics, in particular Smith [1] and Marshall [2]. Writing when Newton's ideas have been generally accepted, the influence of his concept a force as the agent of change is evident in their writing, especially Marshall's. We have relied on the well-known texts by Samuelson [3] and Varian [4] for contemporary treatments.

At the advanced level, the development of economics appears to have moved away from mechanics in favor of a pure mathematical approach. We do not attempt to reconcile these developments in terms of the Newtonian way of thinking, with the possible exception of a short discussion in Section 6 on deBreu's development of general equilibrium, where we rely on the graduate level text [9].

## 2   Definitions

Newton starts his Principia with a chapter wherein he defines the concepts that are required to formulate his laws. In this section, we define the analogous economic concepts. After introducing the economic analog to a particle and a physical body, we present the analogs to the requisite kinematic and dynamic variables, which are summarized in Table 1. We distinguish between two analogies, an impedance analogy where the flow of goods is determined by price, and a mobility analogy where the flow of value is determined by the quantity demanded. Newton's development of mechanics follows the mobility analogy, whereas the theory of demand predominantly follows the impedance analogy. The latter is consistent with hydrodynamics and Maxwell's development of electrodynamics familiar to electrical engineers. We show this in detail in Appendix A.

| | | Mechanics | | Economics | | |
|---|---|---|---|---|---|---|
| | | | | *Mobility* | *Impedance* | |
| *Kinematics* | $q$ | Position | m | Balance | Acquisitions | # |
| | $v$ | Velocity | m/s | Demand Level | Quantity Demanded | #/yr |
| | $a$ | Acceleration | m/s$^2$ | Demand Extension | Needs | #/yr$^2$ |
| *Dynamics* | $p$ | Momentum | Nm | Value | Price | $/# |
| | $F$ | Force | N | Want | Desirability | $/#-yr |

Table 1: Economic analogs to the kinematic and dynamic variables, both in the electrical impedance and the Newtonian mobility analogy.

## 2.1 Inertia as Demand

Newtonian mechanics describes the behavior of an object and the theory of demand and supply describes the behavior of an economic agent. In Newtonian mechanics, the object is idealized as a point particle. We refer to an agent whose behavior is limited to either demanding or supplying as a demander (see Table 2a). Such a demander can equally well represent a supplier or it can even alternate between these two roles. This is unusual in economics, where these roles are assigned to distinct agents.

A particle has inertia[2] and, analogously, a demander has demand (or supply). We intuit that an agent maintains a flow of goods due to its demand and, analogously, a particle maintains a velocity due to its inertia. Mass is a measure of the inertia and the price elasticity of demand is used by economists as a measure of the demand. The concept was first defined by Marshall in [2], who referred to it as the elasticity of wants. According to Marshall,

> *The amount of the commodity demanded may increase much or little according as the demand is elastic or inelastic ...*

We see how Marshall's elasticity $\varepsilon$ is analogous to the inverse of the mass $m$, i.e., $\varepsilon = 1/m$, and, hence, it is the price inelasticity that is properly analogous to the mass (see Table 2a).

| | Mechanics | Economics |
|---|---|---|
| | Particle | Demander |
| | Inertia | Demand |
| $m$ | Mass | Price Inelasticity |
| $\varepsilon = 1/m$ | Inverse Mass | Price Elasticity |

(a) A demander as a point particle.

| | Mechanics | Economics |
|---|---|---|
| | Body | Entity |
| | Total Inertia | Aggregate Demand |
| $m_{\text{TOT}} = \sum m_i$ | Total Mass | Aggregate Inelasticity |
| $\varepsilon_{\text{MUT}} = \sum \varepsilon_i$ | Inverse Reduced Mass | Aggregate Elasticity |

(b) Aggregate demand by summing over agents indexed by $i$.

Table 2: Particles and bodies locate the presence of inertia in mechanics, while demanders and entities locate the presence of demand in economics. Price elasticity is a measure of demand analogous to the inverse of the mass as the measure of inertia.

Marshall's use of elasticity is somewhat unfortunate in the present context since, in engineering, elasticity refers to the resistance of bodies to deformations rather then the inverse of inertia. However, the terminology is firmly established both in economic theory and in business practice. Another unfortunate circumstance is that contemporary economic theorists define the price elasticity of demand in a slightly different manner, i.e., in terms of percentage increases rather than absolute ones. In business practice, Marshall's definition is still the prevalent one.

The analog allows us to aggregate the demand from single demanders to multi-agent systems in the same manner as is done for point particle systems. In this way, general economic entities such as corporate bodies can be considered analogous to physical bodies. The total mass is analogous to the aggregate price inelasticity. The reduced mass is analogous to the aggregate price inelasticity and we refer to its inverse as the mutual price elasticity (See Table 2b).

---

[2]We exclude massless particles like photons here for obvious reasons.

## 2.2 Kinematics as Trade Flow

### 2.2.1 Physical Space as Commodity Space

We refer to the economic analog for physical space as commodity space.[3] Physical space has at most three dimensions, whereas commodity space has a dimension for each distinct type of good. We refer to a coordinate dimension in commodity space as an account. Goods of a common type are said to be fungible and economists refer to them as commodities. This condition allows us to summarize the total amount of a commodity passing hands by a single account balance number, analogous to summarizing total distance travelled with a single coordinate value. Alternatively, within the impedance analogy, we can think of a coordinate value as the total acquisitions, analogous to a total amount of fluid or electrical charge. (See Table 1a.)

Points in physical space are identified using a coordinate system. Further borrowing terminology from accounting, we refer to its analog as a chart of accounts. Figure 1b shows a choice of basis for a commodity plane spanned by a chart consisting of two accounts. Unlike the physical space of Newtonian mechanics, which is considered Euclidean, commodity space is merely a space of points with no structure other than enough smoothness to allow us to do calculus. For instance, the account balances need not share the same units and, hence, the Euclidean metric has no a priori economic meaning since one cannot directly compare apples to oranges.

In practice, the judgement of whether goods are fungible or not to qualify being a commodity depends on their specification. Commodity markets such as the Chicago Mercantile Exchange (CME) provide a list of distinct types of commodities (such as corn or pork bellies) each with a set of contract specifications containing the quality requirements for goods to be considered as such for purposes of trading on the exchange. Also, manufactured products, especially those that are mass produced, are highly fungible and should be considered to be commodities in the current sense. Many other manifestations of commodities occur. Notably, money of a particular type, such as cash or demand deposits, can also be thought of as a commodity.

The choice of a particular chart of accounts is a practical matter, depending on the purposes of the model. One can group all consumption goods in a single basket with a single account on one extreme, or differentiate generically between, say, apples and oranges, or specify precisely the type of apple, its size, color, quality and further attributes on the other extreme. To facilitate the analogy with Newtonian mechanics, we assume the account values to be real numbers, despite that commodities are not infinitely divisible into smaller and smaller units.

### 2.2.2 Motion as Trade

The trade activity of a demander over time can be represented by a time-parametrized path in commodity space, analogous to the representation of the motion of a particle in physical space. Such a path is shown in Figure 1b for two commodities. The activity starts at some initial time $t_i$ and logs all acquisitions and dispositions. This charts a set of balances $q = q(t)$ for each of the commodities at any subsequent time $t$.

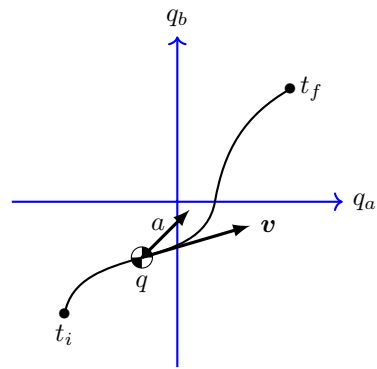

| Mechanics | Economics |
|---|---|
| Motion | Trade |
| Physical Space | Commodity Space |
| Coordinate | Account |
| Coordinate System | Chart of Accounts |

(a) Trade as motion in a commodity space spanned by accounts that serve as coordinates.

(b) Trade activity as the path of the motion in the commodity plane. The quantity demanded $v$ is a vector tangent to the path. Its rate of change is given by the needs vector $a$.

Figure 1: The analogy between motion and trade.

### 2.2.3 Velocity as Quantity Demanded

At any given moment in time, a particle has a specific velocity, equally a demander demands a specific certain quantity of goods. The measure of velocity is the distance traveled per unit of time and the measure of quantity demanded is the amount of a commodity acquired per unit of time. Velocity is defined as the tangent vector to the path of motion. We thus define the quantity demanded as the tangent vector to the path of trade activity. This is shown graphically in Figure 1b for a flat two-dimensional commodity space.

---

[3]Apparently, this terminology was first used by Fisher [10] but has not been adopted in economic texts.

Analytically, this means that the derivative $v := \mathrm{d}q/\mathrm{d}t = \dot{q}$ that represents the quantity demanded is a directional derivative. In this way, the quantity demanded gives the time rate at which the agent is acquiring or disposing of the commodity. Alternatively, and more appropriate to the mobility analog, would be to consider the analog to velocity to be a covector $\mathrm{d}q = v\,\mathrm{d}t$, so that $v$ represents the marginal change in the account level(s) per unit of time. (see Table 1 and Appendix A.)

Economists distinguish between a quantity demanded and one supplied and both are considered to be non-negative numbers. However, our definition of $v$ implies that it is a vectorial quantity, taking on both positive and negative component values. This allows us to unify the two economic concepts into a single vectorial quantity.

### 2.2.4　Acceleration as Needs

When a demander needs more (or less) of a commodity, it increases (or decreases) the quantity demanded. In economics, this is also referred to as an extension (or contraction) of the demand, terminology that is consistent with the mobility analogy. It is analogous to the acceleration (or deceleration) of a particle in mechanics. In the flow picture of the impedance analogy, we refer to the analog to acceleration as the needs of the agent. It is also vector, expressing how much more or less of the mix of commodities an agent needs to demand or supply. (see Table 1.)

### 2.2.5　Center of Mass as Center of Demand

Like a particle, the demander is considered to be highly idealized, only representing a point location of demand with inelasticity $m$ (✺in Figure 1b). For a system of point particles distributed in space, there is a unique point called the center of mass which moves as if it were a point particle with mass $m_{\mathrm{TOT}}$. The analogous center of demand represents an agent whose demand is governed by the aggregate inelasticity of demand (see Table 2b). The kinematics pictured in Figure 1b apply equally well to the behavior of the center of demand for multi-agent economic systems. The economic analogs to the account balances and the quantity demanded vector are shown in Table 3.

|  | Mechanics | Economics |
|---|---|---|
| $q_{\mathrm{CM}} = \sum m_i q_i / m_{\mathrm{TOT}}$ | Center of Mass | Center of Demand |
| $v_{\mathrm{CM}} = \sum m_i v_i / m_{\mathrm{TOT}}$ | Center-of-Mass Velocity | Aggregate Quantity Demanded |
| $a_{\mathrm{CM}} = \sum m_i a_i / m_{\mathrm{TOT}}$ | Center-of-Mass Acceleration | Aggregate Needs |

Table 3: Position and velocity analogs for multi-agent economic systems. The index $i$ ranges over agents.

The concept of a center of demand permits us to treat multi-agent systems as if they were a single agent system —having effective balances, demands, and needs— analogous to the concept of the center of mass in mechanics.

## 2.3　Dynamics and the Price Mechanism

### 2.3.1　Momentum as Price

Newton defines the quantity of motion to be the product of the mass with the velocity, or $p = mv$. This is now referred to as momentum. In economics, if we measure such a quantity for trade in terms economic value, say in \$ per unit of the commodity demanded, then the analogous "value of the trade" in that commodity represents the price the demander ascribes to the commodity (see Table 1). Economists also refer to it as the agent's reservation price, borrowing terminology from the auction process. In business it is also referred to as the transfer price. The engineering diagram in Figure 2b emphasizes the personal nature of the vector $p$ by attaching it to the demander.

There is a momentum associated with each dimension of space. Analogously, there is a price associated with each type of commodity. In economics, the analogous relationship to the definition $p = mv$ of momentum is called a demand schedule (see Table 2a), referring to the use of tables rather than to a function to match price and quantity demanded. Economists view the price as the independent variable, so the relationship is better expressed as $v = \varepsilon p$, where the quantity demanded is the dependent variable and $\varepsilon$ is known is the price elasticity of demand (see also Table 2b). In fact, this is also the case in mechanics, since the causality of the mass is such that velocity can only be changed by changing its momentum by applying a force. The graph of a demand schedule, such as the one shown in Figure 2c, should therefore be read off from the vertical to the horizontal axis.

We exploit our vectorial definition of $v$ to create a unified demand schedule, applicable to an agent acting either as a demander or a supplier, consistent with the definition of momentum. The analogous relationship $v = \epsilon p$ for a

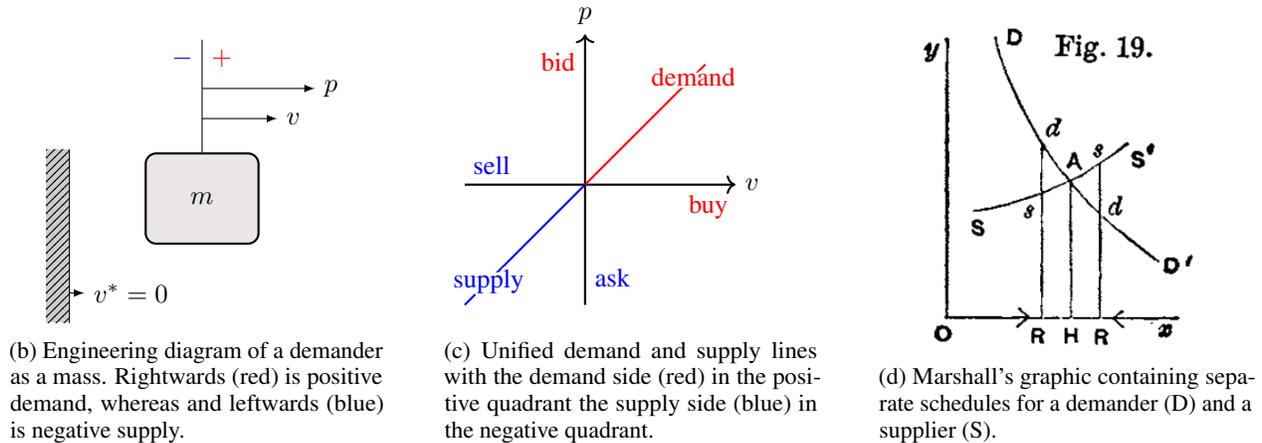

(a) A demand schedule is analogous to the definition of linear momentum.

(b) Engineering diagram of a demander as a mass. Rightwards (red) is positive demand, whereas leftwards (blue) is negative supply.

(c) Unified demand and supply lines with the demand side (red) in the positive quadrant the supply side (blue) in the negative quadrant.

(d) Marshall's graphic containing separate schedules for a demander (D) and a supplier (S).

Figure 2: Unified demand and supply schedules into a single linear vectorial schedule.

demand schedule should be read as a single vector equation for all accounts simultaneously, where price $p$ and quantity demanded $v$ are vectorial quantities and the $\varepsilon$ is a non-negative scalar. This means that $p$ can take on both positive and negative values. The choice $m \geq 0$ implies that $p$ has the same sign as $v$. Therefore, if supply is considered negative, then so are the asking prices and, consequently, the bids are positive (see Figure 2c). When there is no demand, the price is zero.

In the theory of demand economists formulate two schedules; one for an agent such as a consumer, who only demands, and one for an agent such as a producer, who only supplies. Figure 2d shows Marshall's original of the familiar picture of the determination of economic equilibrium in terms of crossing demand and supply curves. Contrary to the unified schedule, prices remain positive, elasticities switch signs, and the curves are shifted from the origin. Economists run the quantity supplied in the opposing direction on the ordinate and our vectorial analysis obviates the need for this by using the sign to distinguish between bid and ask prices. As a consequence, the elasticity can be chosen to remain non-negative.[4] In Section 3.1, we provide the justification for shifting the curves. The economists' picture of economic equilibrium in terms of crossing curves is reformulated in Section 6.3.

## 2.4   Force as Want

A force in mechanics formalizes the intuitive idea of a push or a pull. If we think of pushing for something as wanting or desiring to acquire it, then a such a push is analogous to a *want* or a *desire*. The opposing pull then represents the *incentive* or *effort* inducing us to dispose of it. As vectors, wants and incentive are the negatives of each other and we use the concept of a want to denote any economic force (see Table 1). When emphasizing its force nature, we refer to a "force of desire" or even a "force of wanting."

The terms want and desire appear frequently in economics texts —albeit informally— to indicate the presence of an economic force. The early literature (e.g., Smith [1] and Marshall [2]) tend to use want, while modern texts (e.g., Samuelson [3] and Varian [4]) tend to use desire. In spoken English, a desire is considered stronger than a want and this is also a useful distinction to make for economic forces.

Newton distinguishes between several types of forces whose economic analogs are important for the formulation of the laws in Section 3. We summarize these in Table 4 and discuss them below.

Newton conceives of the inertial force as the tendency for a body to resist a change in its state of motion as measured by momentum. The analogy between inertia and demand (Table 1) immediately suggests that this is analogous to the force of demand, or force of supply for the opposite (see Table 4). Newton defines the inertial force as the rate of change $\dot{p} = \mathrm{d}p/\mathrm{d}t$ of the momentum and, analogously, we let the force of demand equal the rate of change of the agent's price. The force of demand is thus personal to the agent, analogous to how the inertial force is a property of the body.

---

[4]This choice is predicated on the non-negativity of mass in mechanics. Interestingly, it is possible to develop a consistent theory with non-positive masses (see the discussion in Feynman [8]). However, the choice of non-negative masses is firmly entrenched in mechanics, impelling us to make the analogous choice.

| | Mechanics | Economic |
|---|---|---|
| $\dot{p}$ | Inertial Force | Force of Demand |
| $F$ | Motive Force | Economic Motive |

Table 4: Newton's categorization of forces and their economic analogs.

Newton defines the motive force[5] as a force that is exerted on a body to change its state of motion. Its analog as an economic motive is immediate (see Table 4). In [2], Marshall introduces a concept of the "force of an economic motive" as follows:

> It concerns itself chiefly with those desires, aspirations and other affections of human nature, the outward manifestations of which appear as incentives to action in such a form that the force or quantity of the incentives can be estimated and measured with some approach to accuracy ;

Contrary to the inertial force, the motive force is applied externally and is not a property of the body. Analogously, economic motives are exogenous to the agent, rather than personal like the force of demand.

Like their mechanical counterparts, we can distinguish amongst numerous sources for economic motive forces. In Section 4 we consider several that are analogous to common forces in engineering including the economic analog to the gravitational force (constant needs), the spring force (convenience) and the friction force (friction).

## 3   Laws

Newton builds on and justifies his definitions with a chapter —with the same title as this chapter— wherein he posits his three laws of motion. The following three subsections presents the application of his laws to economics, in order of their numbering. Table 5 summarizes the analogs to his laws with laws or basic principles in economics that are both generally known and considered fundamental.

| Law | Mechanics | Economic |
|---|---|---|
| I | Free Motion | Freedom from Want |
| II | Law of Motion | Law of Demand |
| III | Action = Reaction | Demand = Supply |

Table 5: Newton's three laws of motion and their economic analogs.

### 3.1   I: Free Motion as Freedom from Want

Newton's first law stipulates that a particle remains at rest unless a force acts on it. At rest means that the particle maintains a constant velocity. For economics, this means that the agent maintains a constant quantity demanded, i.e., that the vector representing $v$ is constant in both magnitude and direction. In mechanics, the condition is also referred to as free motion, meaning that the motion is free of a net force. Since we interpret an economic force as a want, we formulate the economic analog of this condition as freedom from want. (see Table 5).

Newton's first law requires that the condition of being force-free can be determined independent of the behavior of any other object. Rest in an airplane going at constant speed can be determined with the window shades closed. Analogously, we assume that the condition of freedom from want can be determined by an agent without considering the quantities demanded of any other agents.

Newton's first law is considered a restatement Galileo's principle of relativity, which stipulates that only the relative velocity of two particles with respect to each has physical meaning. The economic analog to the relative velocity is known as the excess demand (Table 6) and the implication of the first law is that only the excess demand between two agents is economically meaningful. It is vector and positive, negative, or zero values indicating whether one agent demands from, supplies to, or does not trade with the other. Hence, an agent may represent a middleman, acting simultaneously as a demander and a supplier to two counterparties.

---

[5] Newton refers to the "impressed force" in the formulation of the definition, while referring later to the motive force. According to Maxwell ([11]), the impressed force corresponds to the impulse of the motive force instead. Indeed, Maxwell refers to the electrodynamic analog of the motive force as the electro-motive force (or EMF).

| | Mechanics | Economics |
|---|---|---|
| $v = v_2 - v_1$ | Relative Velocity | Excess Demand |

Table 6: The excess demand of one demander over another is a vector quantity, analogous to the relative velocity of particles.

In the modern interpretation, the first law postulates both the existence and the equivalence of inertial reference frames. An inertial reference frame is a coordinate frame wherein the first law is valid and, hence, it is moving at a constant velocity $v^*$. For economics, we refer to a demand frame and its existence implies that, in principle, a benchmark level of demand $v^*$ exists that can be used to determine excess demand. In mechanics one chooses a relatively immovable object such as the wall in Figure 2b. We set its velocity $v^* = 0$ and all velocities are then assumed to be relative to this. In economics, a close-to perfectly inelastic entity such as a market consisting of many agents acting collaboratively, can be used to by an agent to mark its demand to the market. We set its quantity demanded $v^* = 0$ and the quantities demanded of the individual agents are then assumed to be the excess over the market.

When modelling, it is convenient to mark all quantities demanded to a single market whose $v^* = 0$, analogous to the engineering practice of choosing a single inertial reference frame. In this way, all the unified demand curves are guaranteed to pass through the origin as required. Shifts in the demand curve are achieved by switching to a different demand frame. The amount of the shift as measured on the demand axis is determined by the relative demand of one demand frame over the other.

### 3.2 II: The Law of Motion as the Law of Demand

In the formulation of his second law of motion, Newton equates[6] the motive force to the change in motion of an object. In the light of the analogs of the previous section, this translates roughly into the statement that the more something is wanted, the higher it is bid up. This coincides with the intent behind the law of demand and, therefore, we consider this law to be the analog to Newton's second law (see 5).

Furthermore, Newton's formulation implies the law concerns vector equations. These are written either as

$$F = \dot{p} \qquad \text{or} \qquad F\,\mathrm{d}t = \mathrm{d}p \qquad (1)$$

corresponding to the mobility and the impedance analogy, respectively. Both equations specify a cause-and-effect relationships, with the cause on the left of the equality sign and the effect on right of it. In mechanics, a force is the *cause* of a change in momentum. For economics, this implies that the agent's motive $F$ *cause* its price change $\dot{p}$.

In the $F = \dot{p}$ formulation, the force $F$ is interpreted as the time rate of a flow $\dot{p}$ of momentum.[7] A want $F$ is interpreted as a flow of economic value that causes the rate of additions $\dot{p}$ to the value $p$ of the commodity. A similar point of view was taken by Adam Smith when formulating his value-added theory of price (see [1]). This is still the basis for pricing in the theory of production, in national accounting, and familiar to many from the manner in which value-added taxes are formulated.

In the $F\,\mathrm{d}t = \mathrm{d}p$ formulation, the impulse of the force $F\,\mathrm{d}t$ drives the change in momentum $\mathrm{d}p$. Here, we interpret the marginal change $\mathrm{d}p$ in price to be caused by the inducement $F\,\mathrm{d}t$ (see Table 7). In economic theory, this approach to price forming is the one taken by the marginalists. We also recognize it in the price discovery process in auctions and the equation can be interpreted as stating that an agent's wants induce it to bid up the price.

| | Mechanics | Economics |
|---|---|---|
| $F\,\mathrm{d}t$ | Impulse of the Force | Inducement |

Table 7: The analogy between the impulse of the force and an economic inducement.

Using a demand schedule, we determine the dynamics of the quantities demanded from the price dynamics. We substitute the unified demand schedule $p = mv$ (see Table 2a) into the statement of the second law $F = \dot{p}$ (Equation 1) to find:

$$F = ma + v\dot{m}. \qquad (2)$$

It follows that a want can have two effects (see Figure 3):

---

[6]Newton actually posits proportionality, but the proportionality constant is without exception taken to be 1.

[7]Newton, in fact, referred to the overdot notation as a fluxion.

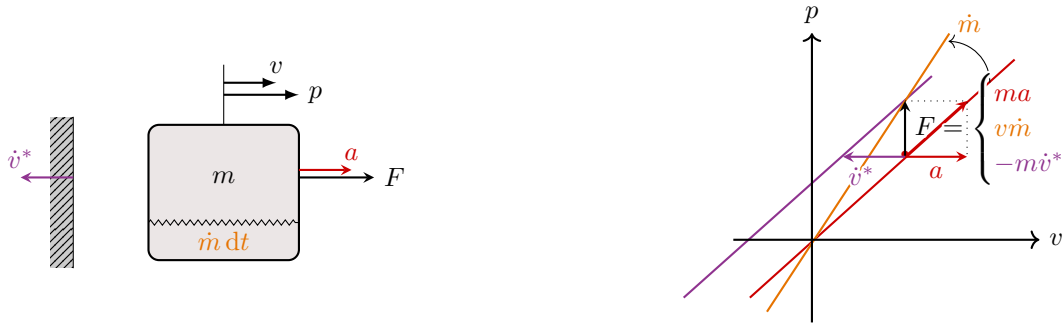

Figure 3: The analogy between Newton's second law and the law of demand. The want $F$ relates to a movement along (red), or a rotation (orange) of the demand curve. A shift in the demand curve (purple) corresponds to a fictitious force or envious desire that arises due to a comparison with an elastic benchmark.

1. $F = ma$: a movement along the demand line, i.e., an extension $a = \dot{v}$ of the demand caused by the agent bidding up the price, and

2. $F = v\dot{m}$: a rotation of the demand line, i.e, an adjustment $\dot{m}$ of the price elasticity of demand that quantifies a change in the agent's personal values.

Another effect should be added to these if we allow for an extension in the demand of the reference market:

3. $F = -m\dot{v}^*$: a shift of the demand line, i.e., a contraction $\dot{v}^*$ of the reference demand leading to a perceived increase in its price (a fictitious economic force).

If we ascertain that the price elasticity and the ambient conditions are relatively constant compared to the changes in demand, then the second law can be restated in the $F = ma$ form and only movements along the demand curve need be considered. This is the typical usage in engineering. In the impedance analogy, this version of the second law states that desirability and needs are proportional to each other. Depending on the elasticity, the same needs may give rise to widely varying desires.

Although economists rarely explicitly consider rotations of the demand curve, they frequently consider convex demand curves rather than the linear ones that are consistent with the definition of momentum (see, e.g., Marshall's curves in Figure 7c). Particles whose mass explicitly depends on the velocity cannot be entertained in classical mechanics and, hence, we exclude convex demand curves from consideration. However, the effect may nevertheless be ascribed to an implicit rotation that occurs due to the timing of changes in the elasticity. For instance, the mass of a rocket ship decreases while it picks up speed because it emits exhaust fumes. The demand curve for the analogous economic process would indeed appear as a convex curve if we did not account for the effects of $\dot{m}$ on the price movements conform Equation 2.

Economists commonly shift the demand curve when the ambient economic conditions change. Such a shift is analogous to an analysis in a non-inertial reference frame. For example, from the perspective of an accelerating train, one is pushed back into one's chair by what appears as a fictitious force. Analogously, if the reference market changes its quantity demanded at a rate $\dot{v}^*$, the agent will resist changing its own quantity demanded as it desires to preserve its initial demand $v$. To force it to keep up with the changing market conditions, an it has to be provided with an incentive $F = -m\dot{v}^*$. Once the market has completed its adjustment, the agent will be left with a price change $\Delta p = \int m\dot{v}^* \, \mathrm{d}t$ that serves to shift its demand curve. If we wish, we may avoid having to shift the demand curve by agreeing on a single reference frame to mark any elastic agent demand to market for the entire analysis. Indeed, from the perspective of the ground, the fictitious force appears as the force of the train on the rider and can be seen as the analog to a movement along the demand curve, rather than a shift.

### 3.3 III: Action=Reaction as Demand=Supply

Newton's third law relates to an interaction between two particles and, consequently, its economic analog must relate to trade between two agents, a demander and a supplier. Newton postulated that this interaction consisted of a pair of equal but opposing forces, an action and an opposing reaction. Analogously, we assume that trade involves two equal but opposing wants, a force of demand and a force of supply. The third law is commonly summarized by that statement that action=reaction and we summarize its economic analog as demand=supply, with the understanding that this concerns the forces of demand and supply (see Table 5).

The idea that price is determined in the context of a transaction between agents is already present in the writing of Adam Smith. In his famous diamond-water paradox [12],

> *A diamond, on the contrary, has scarcely any use-value; but a very great quantity of other goods may frequently be had in exchange for it.*

Smith illustrates the distinction he makes between what he calls value in exchange vs value in use. The former is established through trade between agents and the latter can be determined by the agent itself. The third law implies, therefore, that $p$ corresponds to Smith's value in exchange. This excludes an interpretation such as present value, which qualifies as a value in use. Indeed, a present value calculation can be performed without a counterparty to the transaction and cannot be a price $p$ in this context.

Like the expressions for the second law, the third law is also written in on of two ways:

$$F_s = -F_d \qquad \text{or} \qquad F_s \, \mathrm{d}t = -F_d \, \mathrm{d}t \qquad (3)$$

Here $F_d$ and $F_s$ are the motive forces of the demander and the supplier, respectively. However, the equalities should now be interpreted as a simultaneity, rather than a cause and effect relationship.

In the $F_s = -F_d$ formulation, the third law states that the economic value flows from the demander (the action) to the supplier (the reaction). (see Figure 4a) This implies that no value is lost during trade and, hence, a law of conservation of value in exchange is shown to apply, analogous to the law of conservation of momentum. If we agree that positive flows are said to be debited and negative flows credited (see also Appendix A), then the law can also be interpreted to imply that the total value of debits equals the total value of credits. Therefore, the third law can be seen to enforce a form of double-entry bookkeeping for economic value: the account of the supplier is debited at same rate that the account of the demander is credited.

In the $F_s \, \mathrm{d}t = -F_d \, \mathrm{d}t$ formulation, the third law describes how the increasing marginal cost of the supplier is balanced by the diminishing marginal utility of the demander. In Figure 4b we show how economists' picture of demand and supply implies such a balance consistent with the third law. Because the demand line is downward sloping, the sign change is automatically accounted for.

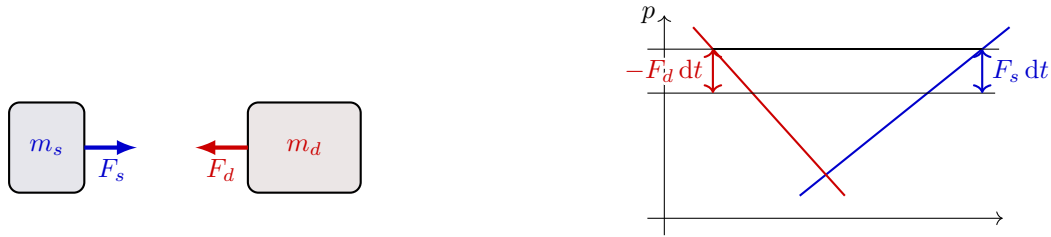

(a) As a conservative flow of value in exchange: the supplier's debits equal the demanders credits.

(b) As a balancing of the diminishing marginal utility of the demander with the increasing marginal cost of the supplier

Figure 4: Newton's third law of motion in the mobility analogy (left) and the impedance analogy (right).

Newton's third law shows how price can be used as a common unit of account, one of the roles of money. Although price itself is personal to each agent, they do agree on the marginal price adjustments during an exchange. To wit, a credit entry $\mathrm{d}p_a$ in commodity account of a party in a transaction must equal the debit entry $\mathrm{d}p_b$ in the corresponding account of the counterparty to comply with the third law. Over time, these entries integrate, thus describing the creation and destruction of credit money that represents the same amount of value to all agents.

In the modern formulation, the third law is extended to systems consisting of multiple particles and is reformulated as the law of conservation of momentum. Such an extension is carried over to economics in a straightforward manner. This leads to an analogous economic law of conservation of value for systems closed to the entry or exit of agents. Alternatively, it leads an overall balancing of credits and debits within an economic system, consistent with the principles of accounting. Finally, when reasoning with price rather than value, it is perhaps more appealing to consider the average per-capita price level in the system. The law then implies the stability of the average price level in a closed economic system.

Conservation of value implies that the demand of the system as a whole depends only on those wants that are exogenous to the system. In fact, the aggregate demand follows the schedule $p_{\text{TOT}} = m_{\text{TOT}} v_{\text{CM}}$ and the second law now applies to the total value $p_{\text{TOT}}$ or, alternatively, to the average price level. The center of demand (see Table 2b) thus behaves as if it were a single agent and this justifies extending the analysis to systems of particles.

In this interpretation, the third law can be seen to apply to aggregate demand. In a closed economic system, endogenous demand is met by endogenous supply, appearing in canceling debit-and-credit pairs of economic value between two parties. This implies that the dynamics of the system as a whole depends only on those wants that are exogenous to the system. In fact, the aggregate demand follows the schedule $p_{\text{TOT}} = m_{\text{TOT}} v_{\text{CM}}$ and the second law now applies to the total $p_{\text{TOT}}$ or, alternatively, to the average price level. It follows, furthermore, that the center of demand (see Table 2b) behaves as if it were a single agent.

## 4 Force Laws as Price Drivers

To make Newton's laws useful in engineering practice, one has to specify the motive force $F$ that drives the change in the momentum $\dot{p}$. Analogously, to make the law of demand useful for economic engineering, one has to specify the economic motive force $F$ that drives the rate of change in price $\dot{p}$. In mechanics, such a force specification is known as a force law or a constitutive relationship. In economics, force laws are also referred to as price- or value drivers.

The simplest, non-trivial price driver is a constant one, driven by the basic needs, say $g$, of an agent. Adam Smith argued [12] that although the needs of all agents are limited and essentially identical, their desires may differ substantially:

> *The capacity of [the landlord's] stomach bears no proportion to the immensity of his desires, and will receive no more than that of the meanest peasant.*

We translate this into a familiar mechanical setting by considering these needs as analogous to a gravitational acceleration $g$. Then the corresponding desire $F_g = mg$ depends on the elasticity of the particular agent. With constant needs, desires are proportional to price inelasticity and the desires of highly inelastic agents (Smith's landlords) will be appear immense (see Equation 2).

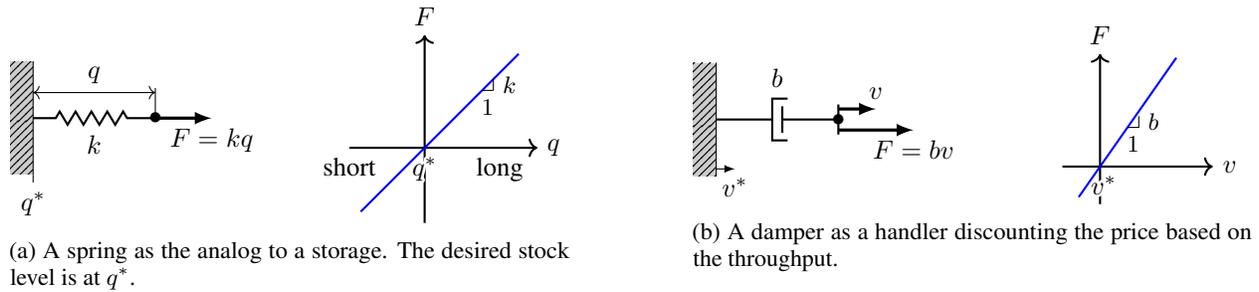

(a) A spring as the analog to a storage. The desired stock level is at $q^*$.

(b) A damper as a handler discounting the price based on the throughput.

|  |  | **Mechanics** | **Economics** |
|---|---|---|---|
| *Law* | $F = kq$ | Spring | Storage |
| *Force* | $F$ | Spring | Convenience |
| *Variable* | $q$ | Extension | Stock |
| *Parameter* | $k$ | Stiffness | Stock Inelast. |

|  |  | **Mechanics** | **Economics** |
|---|---|---|---|
| *Law* | $F = bv$ | Damping | Friction |
| *Force* | $F$ | Damping | Friction |
| *Variable* | $v$ | Velocity | Throughput |
| *Parameter* | $k$ | Coefficient | |

Figure 5: Force laws and their economic analogs and the naming of the corresponding forces, driving variables, and parameters.

*Storage Laws:*

Scarcity is traditionally seen as an important price driver in economics. In the words of Adam Smith [1]

> *If the commodity be scarce, the price is raised.*

This can be modelled mechanically by a potential force, such as is provided by a spring. The force law for a linear spring is given in Table 5. In the theory of storage, the balance $q$ is thought of as the inventory stock of the commodity. If the inventory level is low and the commodity is scarce, the position is said to be short and, when high and abundant, it is said to be long. Notice how suggestive this type of language is of the physical shortening and lengthening of a spring.

The stock quantity $q$ drives the price with a force $F = kq$ that expresses the convenience (or inconvenience) of holding that quantity of the commodity. It is also known as the convenience yield [13] or the own yield [14] of the commodity.

Like the spring force, the force of convenience is a restoring force, so $F$ incentivizes the agent to reduce stocks when inventories are high and increase stocks when they are low. At the desired level $q = 0$ and $F = 0$ (See Figure 5a.) The analog to the spring stiffness is the stock elasticity (of the convenience), not to be confused with the price inelasticity (of demand), the analog to the inertia.

Non-linear storage laws are more common than not, both in economics and in mechanics. For example, an inventory stockout is analogous to the spring in Figure 5a hitting the wall. The arbitrarily large inconvenience force from the limits of the storage are analogous to the arbitrarily large motive force from the wall. In addition, if the spring is stretched beyond its elastic region into the plastic region, its stiffness $k$ decreases and eventually vanishes entirely upon fracture. An economic instance of this phenomenon, when the commodity concerns money, is a liquidity trap, where additional money stocks fail to provide any additional convenience. For a final example, we extend the gravitational analog for constant needs given above to arbitrarily large distances or stock amounts. The analog to Newton's law of universal gravitation then implies that the convenience $F \propto m/q$ and, hence, that the Smith's universal needs $g \propto 1/q$ are inversely related stock level, vanishing entirely in the limit when $q \to \infty$.

*Friction Laws:*

Trade friction is also a ubiquitous price driver in economics. Economists also refer to any process that impedes the flow of goods as friction. This is entirely analogous to usage in mechanics, where one thinks of friction as resisting the motion. In practice, one imagines a handler such as a broker, a shipper, or other such intermediary who charges a cut $F = F(v)$ depending on the flow $v$ of goods being handled.

The constitutive law for a mechanical damper can serve as a linear model for trade friction (see Table 5 and Figure 5b). Its specification implies that the impeding force $F = bv$ is always in the direction of the quantity demanded, thus guaranteeing that the handler has to be incentivized to provide its services. Therefore, any law that complies with this requirement may serve as a friction law, as well as a linear one. For example, handling costs can be for fixed fee $F = b \operatorname{sgn} v$, analogous to static friction, perhaps maintained by a constant fee, which is analogous to kinetic friction. Alternatively, at high velocities, the linear viscous friction becomes quadratic in velocity in accordance with the drag equation.

## 5 Mechanical Energy as the Economic Surplus

In mechanics, Newton's definition of a force is used to define the concepts of work and energy. We proceed in an analogous fashion, using the Newtonian economic force to develop the concepts of economic benefit and surplus. These are presented in the following two subsections and summarized in Table 6a.

|  | Mechanics |  | Economics |  |
|---|---|---|---|---|
| $\Delta W = \int \dot{W} \, \mathrm{d}t$ | Work | J | Benefit | \$/yr |
| $\Delta Q = \int P \, \mathrm{d}t$ | Heat |  | Consumption |  |
|  | *Energy* |  | *Surplus* |  |
| $T = \frac{1}{2m} p^2$ | Kinetic |  | Direct |  |
| $V = \frac{1}{2} k q^2$ | Indirect |  | Potential |  |
| $H = T + V$ | Mechanical |  | Disposable |  |

(a) Economic benefit as the analog to mechanical work energy and the various forms of energy and their analogous types of economic surplus.

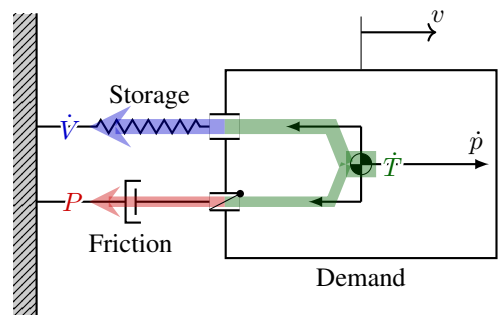

(b) Allocative efficiency as a conservative flow of economic surplus. The check valve indicates the unidirectional character of the distribution and diversification of the surplus.

|  | Power Balance | Economic Efficiency |
|---|---|---|
| $\dot{T} + \dot{V} = 0$ | Reactive | Reallocative |
| $\dot{T} + P = 0$ | Active | Consumptive |
| $\dot{T} + \dot{V} + P + \dot{W} = 0$ | Total | General Allocative |

(c) Power balances as allocative efficiencies.

Figure 6: Economic surplus and its flow as analogs to energy and power.

## 5.1 Work as Benefit

In mechanics, a force does work when an object is moved over some distance. Analogously, we say in economics that a want allocates benefits when an agent acquires some quantity of goods. The rate at which this is done,

$$\dot{W} = Fv \tag{4}$$

is called the delivered power in mechanics. In economics, we refer to it as the benefit allocation rate.

The product should be read as an inner product between the wants and the quantity vectors, thus yielding a scalar. In mechanics, the positive direction is customarily taken from the supply of power to the load. In economics, following the same custom, we take the allocation rate to be positive from the benefactor to the beneficiary.

Economically, the inner product operator represents the process of attaining satisfaction. Due to its symmetry, the operator can be thought of in two ways: either the want $F$ is *satisfied* by the quantity demand $v$, or the the want is *met* by the flow $v$.[8] For the special case when $F$ is the force of the needs $g$ (see Section 4) and the power is given by $\dot{W} = gmv$ we say that the needs $g$ are fulfilled.

The total benefits $\Delta W$ received over a period of time are determined by integrating the allocation rate $\dot{W}$ over the actual trade path in the commodity space, such as the one in picture in Figure 1. The integral in Table 6a should be interpreted as a line integral along the trade path. In general, there is no guarantee that a different path connecting the same initial and final account balances accrues the same benefits. In Section 5.2.2 we categorize the cases where accrued benefits are path independent and only depend on the account balance.

## 5.2 Energy as Surplus

### 5.2.1 Kinetic Energy and the Direct Surplus

Marshall defined economic surplus (which he referred to as economic rent [Figure 7c]) for consumers and producers in terms of areas in a supply and demand graph (see Figure 7b). Using the concept of a unified demand schedule, a single notion of surplus – applicable to demanders and suppliers alike, – emerges and we immediately recognize the mechanical analog to be kinetic energy (see Figure 7a).

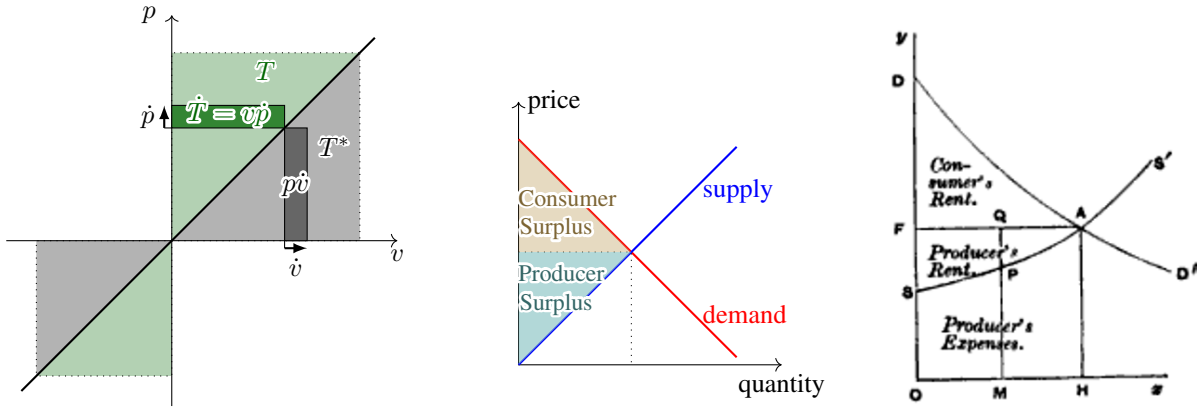

(a) Economic surplus and economic costs as the economic analogs to kinetic energy $T$ (green) and coenergy $T^*$ (gray) on a unified demand schedule.

(b) Typical economics graphic of the economic surplus separating consumer from producer surplus.

(c) Marshall's original graphic indicating both surplus (rent) and economic costs (expenses).

Figure 7: Economic surplus as analog to kinetic energy.

Economists also refer to the surplus as the gross benefit associated with a good (see [4]). Using the analog to the work-energy principle —or, analogously, the benefit-surplus principle— we can turn this into an equality between benefits and surplus. Starting with the benefit allocation rate $\dot{W}$ we arrive at the surplus allocation rate $\dot{T}$ by reasoning as follows:

$$\dot{W} = Fv = v\dot{p} = \dot{T}. \tag{5}$$

The critical step is formed by the third equality where we use the Newtonian law of demand to express the want in terms of a price movement. The inner product $v\dot{p}$ describes the process at which the agent's price adjustments $\dot{p}$ serve to satisfy its demand $v$ for the commodity. To emphasize this link to trade, we consider $T$ to be the *direct* economic surplus. It is analogous to kinetic energy, so named to emphasize the link with motion.

---

[8]The language of differential geometry gives us the most compelling formulation of the allocation rate. The product in Equation 4 represents the natural action of a 1-form on a vector (see Appendix A). With $F$ a vector, we say that the want is *satisfied* by the quantity $v$. When $F$ is a covector, we say rather that it is *met* by the flow $v$.

We accrue the surplus allocation rate by integrating it over the applicable time period. Because $v\dot{p}\,\mathrm{d}t = v\,\mathrm{d}p$ can be recognized as the marginal increase in surplus due to a price margin $\mathrm{d}p$ over the reservation price, the integration depends only on the final price and we find the expression in Table 6a. The integration is shown graphically in Figure 7a. The area representing the surplus is swept out by the reallocation rate $\dot{T}$ as it moves along the price axis. We see that the surplus is naturally a function of price $p$, analogous to kinetic energy as a function of momentum.[9]

*Kinetic Coenergy as Economic Costs*

Marshall referred to the area underneath the supply line as expenses (see Figure 7c). We refer to this as economic costs, emphasizing that this is a flow of economic value rather than an accounting term for a money flow. The unified picture in Figure 7a shows that it is analogous to kinetic coenergy. (See Table 8.)

|  | Mechanics | Economics |
|---|---|---|
| $T^* = \dfrac{1}{2}mv^2$ | Kinetic Coenergy | Direct Costs |
| $T = pv - T^*$ | Legendre Transform | Surplus = Revenue - Costs |

Table 8: Kinetic coenergy as the direct economic costs and its relationship with the economic surplus.

The economic costs to the agent for extending its demand at a rate $\dot{v}$ are obtained after multiplication by the value $p$ to obtain the rate $p\dot{v}$. Economists refer to $p\dot{v} = p\,\mathrm{d}v$ as an increase in marginal costs. Integrating the former rate over a time period, or the latter marginal up to the current level of demand, gives the agent's economic costs (see Table 8). This is shown graphically in Figure 7a. The economic costs are naturally a function of the quantity demanded $v$.

The nature of economic surplus is further understood by contrasting it with costs. Consider the economic revenue $pv$ to the agent. In Figure 7a it is evident that the area representing revenue is the sum of the areas of surplus (green) and costs (gray). Economically, this means that surplus equals the revenue net of cost (see Table 8) and, hence, can be thought of as a form of economic profit. The surplus becomes a measure of the value an agent receives over and above the intrinsic utility of a commodity to said agent. In mechanics, it is the Legendre transform that gives the analogous relationship between energy and coenergy. The Legendre transform formally establishes a relationship between energy and coenergy as functions (of $p$ and $v$, respectively) rather than as numbers.

### 5.2.2 Potential Energy as Indirect Surplus

In mechanics, when the work done by a force only depends on the change in position and not on the path of the motion connecting these, the force is said to be conservative. Analogously, we speak of a convenience force when benefits depend only on the change in the amount in storage and not on the trade path (see Section 4 for the linear case). In mechanics, the work received is then referred to as potential energy. We refer to its economic analog as indirect surplus.

The direct surplus $T$ can be reallocated to indirect surplus $V$ and back again. This is analogous to kinetic energy being transformed into potential energy and vice versa. This is also known as reactive power in electrodynamics or reversible work in thermodynamics. According to the benefit-surplus principle, the reallocation rate is:

$$\dot{T} = v\dot{p} = -F\dot{q} = -\dot{V}. \tag{6}$$

The critical equality is the third one. Using Newton's second law in reverse, we set the price adjustment $\dot{p}$ equal to the negative of the convenience force $F = F(q)$. The minus sign indicates that the convenience force, like the spring force, is assumed to be a restoring force and acts in the direction inverse to the stock $q$ (see Section 4). Simultaneously, the quantity demanded $v$ draws from the inventory at a rate $\dot{q}$. As a result, indirect surplus grows at the expense of direct surplus and vice versa. Hence, if the demander is the beneficiary, the storage becomes the benefactor and vice versa.

To determine indirect surplus, we accrue its allocation rate $\dot{V}$. We recognize $\mathrm{d}V = F\dot{q}\,\mathrm{d}t = F\,\mathrm{d}q$ as the marginal benefit of holding an additional marginal amount $\mathrm{d}q$ at convenience $F$. A storage law determines the convenience $F = F(q)$ as a function of the amount $q$ in storage and, hence, the $\mathrm{d}V$ can be integrated to yield the indirect surplus $V = V(q)$ as a function of the stock $q$. The most elementary storage law is a constant one, in which we find that $V = mgq$, or that indirect surplus increases linearly with inventory level $q$, analogous to potential energy in a gravitational field. The linear storage law from Section 4 gives an expression that depends quadratically on $q$ (see Table 6a), recognizable as the expression of the potential energy of a spring.

---

[9]Economists typically integrate along the quantity axis instead (see, e.g., the distance PQ in Marshall's graphic in Figure 7c) and consider the surplus a number rather than a function. Although the value of the surplus will be the same, the functional dependence on the price is lost and, with it, the dynamics of the surplus allocation.

### 5.3 Dissipation as Consumption

In mechanics, friction impedes motion and in economics friction impedes the flow of trade. In both cases, this implies that —as vectorial quantities— the friction force $F = F(v)$ is in the same direction as the movement $v$. Applying the benefit-surplus principle, we find that:

$$\dot{T} = v\dot{p} = -Fv := -P \leq 0 \tag{7}$$

The minus sign is required to assure that the friction does impede the trade flow. The quantity $P = Fv$ is known as active power or the rate of energy dissipation. The alignment of $F$ and $v$ implies that it is non-negative and, hence, Equation 7 implies that surplus, like energy, exclusively decreases in the presence of friction. In economics, we refer to this process as consumption and characterize an allocation of surplus to friction as consumptive (see Table 6c).

The linear law $F = bv$ leads to quadratic consumptive allocation rates $P = bv^2$ that are, in fact, positive for any non-zero coefficient $b$. Many non-linear laws are applicable, depending on the circumstances. Fixed fee structures of the form $F = b\,\mathrm{sgn}(v)$ lead to rates of the form $P = b|v|$, also non-negative. In Section 6, we analyze quadratic friction analogous to drag forces.

Integrating consumptive allocation over time yields the consumption $\Delta Q$ (see Table 6a). The $\Delta$ notation emphasizes that, like benefits and unlike surplus reallocations, the integral cannot be evaluated in a path-independent manner and, hence, consumption is not a bonafide function.

### 5.4 Power Balance as Efficient Allocation

The analog between energy and surplus implies that, like energy, economic surplus is a conserved property of a closed system. In mechanics, the manifestation of this is that energy can be neither created nor destroyed, but must flow from one body to another. The flow of energy is also called power and a power balance guarantees that it is conservative. For economics, we refer instead to an allocation of surplus and say that the allocation is efficient to indicate it is conservative (see Table 6c).

*Reallocative Efficiency*

When energy is transferred between kinetic and potential forms, the transfer is reversible and the power is called reactive to emphasize that it can flow in either direction. We refer to the analogous allocation between direct and indirect surplus as reallocative to emphasize that the roles of beneficiary and benefactor may switch. Although in Figure 6b, the storage is the beneficiary, this is simply because $\dot{V} = F\dot{q}$ happens to be positive because the convenience and the inventory fill rate vectors are aligned. The reallocative efficiency $\dot{T} + \dot{V} = 0$ that is analogous to the reactive power balance (see Table 6c) follows from Equation 6. We can see this graphically since the rate of reduction of the area corresponding to the direct surplus in Figure 8a equals the rate of increase of the area corresponding to the indirect surplus in Figure 8b.

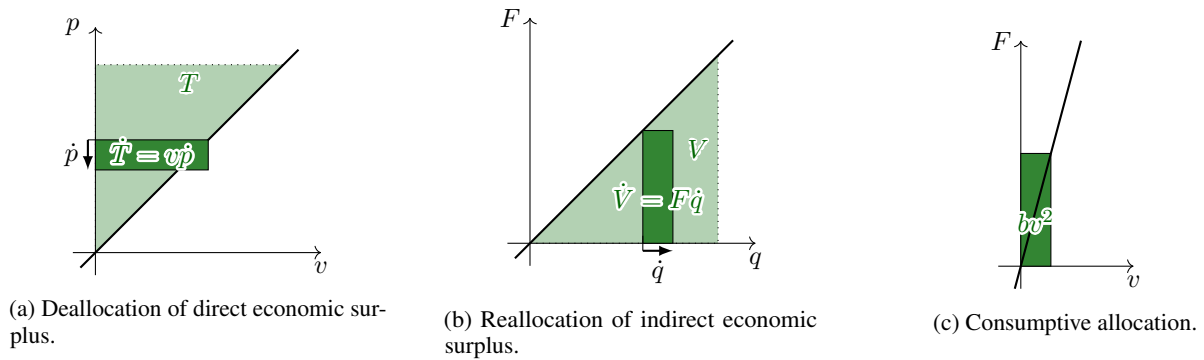

(a) Deallocation of direct economic surplus.

(b) Reallocation of indirect economic surplus.

(c) Consumptive allocation.

Figure 8: Allocative efficiency and the conservation of total economic surplus. The consumptive allocation is necessarily positive.

Integrating the reallocative balance (Equation 6), we find that $H = T + V$ is a constant. In mechanics, the analogous constant is known as the total mechanical energy. We refer to its economic analog as the disposable surplus because it can be reallocated at will, analogous to how the mechanical energy can be converted to any other form of energy.

*Consumptive Efficiency*

When kinetic energy is dissipated by a damper, the transfer is irreversible and the power is called active to emphasize that the energy flow is in one direction only. We refer to the analogous allocation as consumptive to emphasize that the frictional entity is at all times the beneficiary. In Figure 6b, this is suggested by the hydraulic check valve. It is seen graphically by realizing that the area in Figure 8c is at all times positive and, hence, the direct surplus is necessarily deallocated in Figure 8a. The consumptive balance $\dot{T} + P = 0$ follows from Equation 7 and is the analog to the active power balance (see Table 6c). Integrating the consumptive balance, we find that $T + \Delta Q$ is a constant.

In mechanics, the lost energy $\Delta Q$ is thought of as being *dissipated* and the rate of dissipation $P$ is referred to as the active power (see Figure 6b). The picture of dissipation from statistical mechanics is that of the available energy is distributed over the multitude of particles and degrees of freedom. In economics, we picture the surplus as being *distributed* over agents and *diversified* over the different types of commodities.[10] Adam Smith [12] famously imagined an invisible hand that governs economic processes. He writes:

> They are led by an invisible hand to make nearly the same distribution of the necessaries of life, which
> would have been made, had the earth been divided into equal portions among all its inhabitants...

The analogy with statistical mechanics becomes apparent if we think of "they" as agents and "necessaries" as economic surplus as it is diversified over various products and commodity types.

A fundamental law of physics (the second law of thermodynamics), postulates that dissipation is an irreversible process. Extending this to economics, this implies that consumption is also irreversible. This further implies that consumptive allocation, like active power, cannot be negative, as witnessed by the expression $P = bv^2$ for linear friction. Smith's invisible hand can thus be visualized as the thermodynamic arrow of time, irrevocably pushing the system to an equilibrium where surplus vanishes and trade stops.

*General Allocative Efficiency*

The overall power balance requires that the energy flows add up to zero. This corresponds to the general allocative efficiency listed in the top row of Table 6c. This can be accrued over time by integrating to obtain the constancy of total surplus:

$$H + \Delta Q + \Delta W = E.$$

In mechanics, the constant $E$ is known as the total energy and the equality as the law of conservation of energy. An analogous equality in economics occurs for the calculation of GDP as the sum of consumption, investment and government expenditures, customarily written as $Y = C + I + G$. If we think in terms of value rather than money, then if consumption is $\Delta Q$, investment is the indirect surplus $V$, and the government the benefits $\Delta W$, the calculation of GDP is consistent with the law of conservation of surplus.[11]

Smith argued that the invisible hand irrevocably leads markets to attain equilibrium as long as there are no government interventions. Therefore, if we let $\Delta W = 0$ to exclude any interventions, it follows that when the consumption $\Delta Q$ grows in time with $P \geq 0$, the disposable surplus $H$ must decrease and ultimately vanish altogether. Then, any trade activity ceases and the system thus attains equilibrium.

## 6 Equilibrium

### 6.1 Economic Equilibrium as Mechanical Equilibrium

In [2], Marshall introduces the concept of economic equilibrium as follows:

> The simplest case of balance or equilibrium between desire and effort is found when a person satisfies
> one of his wants.

Our definition of an economic force puts us in a position to make the analogy with mechanical equilibrium. In engineering practice, equilibrium is determined with the aid of a free-body diagram. The agent's desires (or wants) and efforts (or incentives) are balanced by adding them vectorially to determine a net want $F_{\text{NET}}$ (see Figure 9). Using our Newtonian law of demand (Equations 1), any net want results in a price adjustment $\dot{p} = F_{\text{NET}}$. Hence, only when the net want $F_{\text{NET}} = 0$ do we obtain a constant equilibrium price. This condition is known as price equilibrium in economics and dynamical equilibrium in mechanics (see Table 9a).

---

[10]For the damper in Figure 6b that represents the handling services, the analogy can be made explicit as follows. The work done on the damper serves to increase the average kinetic energy of the particles that make up viscous fluid it contains. The energy content of the work is thus dissipated over numerous particles, each travelling at its own speed in various directions. For economics, we think of the fluid particles as the, presumably numerous, agents involved in the intermediation of trade flow and the the dissipation process describes how the surplus is distributed among the agents who diversify their portfolios over the available commodity types.

[11]Exports net of imports also makes up part of $\Delta W$. Our analysis does suggest adding the value of the direct economic surplus to GDP as well.

|  | Mechanics | Economics |
|---|---|---|
| $\dot{p} = 0$ | Dynamic Equilibrium | Price Equilibrium |
| $\dot{p} = 0$ & $\dot{q} = 0$ | Static Equilibrium | Competitive Equilibrium |

(a) Types of mechanical equilibria and their economic analogs.

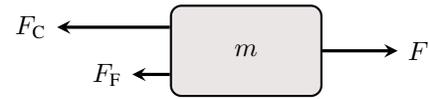

(b) Determining the net want with a free body diagram: $F_{\text{NET}} = F_{\text{C}} + F_{\text{F}} + F$. In the figure $F_{\text{NET}} = 0$ then the agent is in price equilibrium.

Figure 9: Mechanical analysis of economic equilibrium.

Price equilibrium alone does not imply that stocks remain constant since these grow linearly with the quantity demanded. In the special case where $\dot{q} = 0$, i.e. the level of demand equals that of the benchmark market level, inventory stocks also remain constant. In mechanics, this is referred to as static equilibrium and economists refer to competitive equilibrium in this context (see Table 9a).

## 6.2 Stable Equilibrium and the Invisible Hand

An economic equilibrium need not be stable. Adam Smith [1] referred to an equilibrium price value that is stable as the "natural price":

> The natural price is as it were the central price to which the prices of all commodities are continually gravitating. Different accidents may sometimes keep them suspended a good deal above it, and sometimes force them down even somewhat below it. But whatever may be the obstacles which hinder them from settling in this center of repose and continuance, they are constantly tending towards it.

In [2], Marshall makes an explicit analogy with the manner in which stability is described in mechanics:

> When demand and supply are in stable equilibrium, if any accident should move ... from its equilibrium position, there will be instantly brought into play forces tending to push it back to that position; just as, if a stone hanging by a string is displaced from its equilibrium position, the force of gravity will at once tend to bring it back to its equilibrium position.

We formalize these ideas using the analysis of equilibrium in mechanics. There, a distinction is made between stable and asymptotically stable systems. For economics, this implies that an equilibrium is *stable* when the neither the price nor the stock diverge indefinitely from their equilibrium values. The equilibrium is *asymptotically stable* if these values actually converge in the *long run* to their equilibrium values. For instance, analogous to the momentum of a stone hanging from a string, the price may oscillate indefinitely around its equilibrium value when there is no trade friction. In the presence of friction, however, the price fluctuations around the equilibrium value diminish and price and stock will move arbitrarily closely to their equilibrium values. In Section 7, we work this out for several applications.

The conditions for asymptotic stability are precisely those of the invisible hand (see Section 5.4). A friction force implies a consumption allocation $P > 0$ and, as long as no benefits are being allocated and $\dot{W} = 0$, this implies that the disposable surplus shrinks until it vanishes entirely. In the language of dynamical-systems theory, this means that the disposable surplus can be used as a Lyapunov function to prove asymptotic stability, thus formalizing the Smith's picture of the invisible hand as the driver for long-term economic equilibrium to the "natural price."

## 6.3 Mutual Equilibrium as the Two-Body Problem

In [2], Marshall describes the *mutual* equilibrium between two agents, a demander and a supplier, as follows:

> When the demand price is equal to the supply price, the amount produced has no tendency either to be increased or to be diminished ; it is in equilibrium.

Mutual equilibrium is traditionally pictured to the point where demand and supply lines intersect. In Marshall's original picture in Figure 7c, this is the point labeled A and in Figure 10b, this is the point labeled as the origin $(0, 0)$.

The analogous analysis in mechanics is known as the two-body problem, and we show how the economists' picture can be constructed using its solution (Figure 10). In the two-body problem, there is no body that is massive enough to serve as an inertial reference frame. Instead, the inertial frame is attached to the center of mass. Analogously, if there are no perfectly inelastic agents to mark the demand to market, we can we use the pair's center of demand as a benchmark. Since there are no third parties involved, the third law (Section 3.3) implies that the midpoint price $p_{\text{CM}}/2$ is a conserved quantity and we choose this and its corresponding aggregate demand level as the origin of the unified demand lines. The

| | |
|---|---|
| $p = p_d - p_s$ | Bid-Ask Spread |
| $v = v_d - v_s$ | Excess Demand |
| $\varepsilon = \varepsilon_d + \varepsilon_s$ | Mutual Elasticity |
| $v = \varepsilon\, p$ | Marshallian Demand Function |

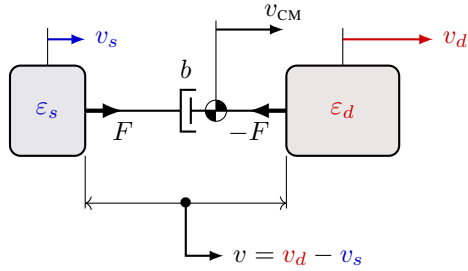

(a) The two-body problem: Two demanders trading through a handler. The agents achieve mutual equilibrium when their quantities demanded equal that of the center of demand.

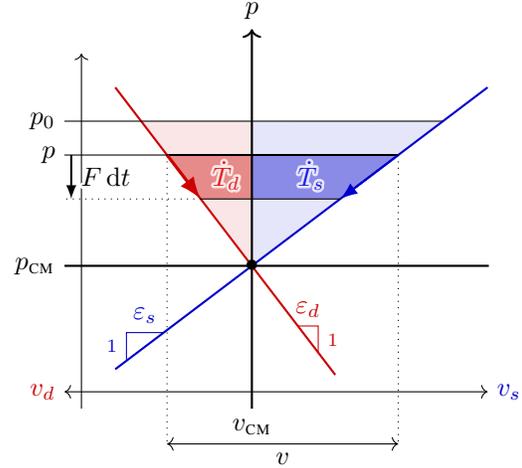

(b) The price dynamics pictured as movements on the familiar economic supply (in blue) and demand (in red) graphic. The disposable surplus is indicated by the areas in light colors and the consumptive allocation at price level $p$ by the parts in the corresponding darker tones.

Figure 10: Economic equilibrium and the two-body problem.

individual agents' prices then represent the spread over the midpoint price and their quantities demanded or supplied the excess over the aggregate level of demand.

Using the center of demand as the benchmark, the two-body problem is reduced to that of a single agent in a force field. In mechanics, the reduced mass is used to do this, but for economics the arguments become particularly intuitive and straightforward by using the aggregate elasticity $\varepsilon = \varepsilon_d + \varepsilon_s$ instead (see Table 2b). It is readily verified that the agents follow a single mutual demand line $v = \varepsilon p$ that relates the mutual excess demand $v = v_d - v_s$ to their mutual price spread $p = p_d - p_s$. In economics, this relationship is known as the Marshallian demand function. It generalizes the single-agent case, which is recovered when one of the agents is perfectly inelastic. For an inelastic supplier, we substitute $\varepsilon_s = 0$ and mutual elasticity $\varepsilon = \varepsilon_d$ reduces to that of the demander alone. The center of demand then coincides with the supplier whose supply line becomes vertical and we recover the situation pictured in Figure 2.

To investigate the forces that push the agents to an equilibrium, we assume the agents' mutual trade is brokered by a handler with friction coefficient $b > 0$. Consistent with the third law, the force of demand must equal the force of supply. In the diagram in Figure 10a this is shown as a flow $F$ of value from the demander, through the handler, to the supplier. In the economists' picture in Figure 10, we show this as equal but opposite inducements $F\,\mathrm{d}t$ for the agents to adjust their prices. The price cut $F$ the handler charges to mediate the trades acts to reduce the price spread over time until it effectively vanishes.[12] In Figure 10a, this manifests itself in time, as the masses representing the agents ultimately move with the same velocity (the quantity demanded) and all trades stops. In Figure 10 it gives rise to movements along the demand and supply curves leading to an equilibrium at the crossing of the two.

The passage to equilibrium can also be argued using the invisible hand alone. In Figure 10, we picture the total surplus $T = T_d + T_s$ by lightly colored areas. The consumptive allocation $P = \dot{T}_d + \dot{T}_s$ corresponds to the portion that is lost over time and we shade this in the darker tones. As long as the trade is not friction free, the consumptive allocation is always positive. Therefore, the areas representing the surplus are reduced until the lack of available surplus causes all trading activity to stop.

Without consumptive allocation, agents would eternally remain out of equilibrium. For instance, if the damper in Figure 10a is replace by a spring that represents some mutual storage facility, the agents would continually stock up or draw from inventory to provide the demand and supply to the counterparty, who would be doing the same thing, albeit $180°$ out of phase. As a result, trading activity would sweep out the areas corresponding to the surplus in both directions and on both sides of the equilibrium point in Figure 10b, never reaching the equilibrium state (see, further, Section 7.2). In practice, this is rare since some form of surplus loss, either through handling, custody, or other types of consumption, is inevitable. The appearance of the invisible hand thus guarantees that equilibrium is reached in the long run.

---

[12]In fact, since the cut $F = \varepsilon bp$, we have that $\dot{p} = -\varepsilon bp$ and the price development consists of an exponential decay at rate $\varepsilon b$ (see also Section 7).

### 6.4 General Equilibrium and the Inertia Tensor

Marshall assumed that the demand for a particular commodity is independent of the demand for any other of the available commodities in the analysis. This assumption is known in economics as partial or Marshallian equilibrium. It is to be contrasted with general or Walrasian equilibrium, which considers all possible types of commodities and their interdependencies simultaneously. In this section, we extend the Newtonian approach to general equilibrium analysis.

The vectorial nature of quantity demanded $v$ and price $p$ allows us to interpret the demand schedule $v = \varepsilon p$ as a linear map $\varepsilon : p \mapsto v$ between vector spaces. In economics, this map is known as the Walrasian demand function. The linear operator $\varepsilon$ is tensor that we refer to as the elasticity tensor. In the one-dimensional case, this tensor reverts to a scalar used in the previous sections.

Using a chart of accounts, we can represent the elasticity tensor as a matrix. It right-multiplies a column vector of prices to give a column vector of quantities demanded. We refer to the diagonal entries of the matrix as principal elasticities and the off-diagonal entries as cross elasticities. In an engineering diagram, the choice of a chart is visualized as a choice of coordinate frame of account directions. In Figure 11, this is shown for a chart consisting of commodities, $a$ (apples) and $b$ (bananas), and a chart for baskets (fruit salads), $\alpha$ and $\beta$.

In the commodity chart, the cross elasticities $\varepsilon_{ab}$ are non-zero (Figure 11a), indicating the presence of the substitution effect. In the particular basket chart we selected, the cross elasticities vanish and the matrix is diagonal. The engineering diagram, the basket accounts are orthogonal, whereas the commodity accounts are not (Figure 11b). Algebraically, the basket accounts form an eigenvector basis for the elasticity tensor, with eigenvalues equal to the principal elasticities of the diagonal matrix. In such an eigen-chart, the Walrasian demand function reduces to the Marshallian demand function and the general equilibrium analysis is converted to a set of independent partial equilibrium problems.

$$v = \qquad \varepsilon \qquad p$$
$$\begin{bmatrix} v_a \\ v_b \end{bmatrix} = \begin{bmatrix} \varepsilon_{aa} & \varepsilon_{ab} \\ \varepsilon_{ab} & \varepsilon_{bb} \end{bmatrix} \begin{bmatrix} p_a \\ p_b \end{bmatrix}$$
$$\begin{bmatrix} v_\alpha \\ v_\beta \end{bmatrix} = \begin{bmatrix} \varepsilon_{\alpha\alpha} & 0 \\ 0 & \varepsilon_{\beta\beta} \end{bmatrix} \begin{bmatrix} p_\alpha \\ p_\beta \end{bmatrix}$$

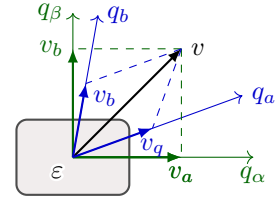

(a) The Walrasian demand function and the corresponding Marshallian demand function in the eigen-basket charts of accounts.

(b) Agent with non-orthogonal commodity accounts (blue) and the corresponding orthogonal baskets (green).

Figure 11: The demand tensor of general equilibrium as an inertia tensor.

The economic surplus is a scalar, independent of the choice of coordinate chart. This is analogous to kinetic energy in higher dimensions. To determine it, we generalize the expression in Table 6a by thinking of the elasticity tensor as a bilinear map $\varepsilon : (p, p) \mapsto T$, taking two copies of the price vector and yielding a scalar value of the surplus. In the matrix representation, this can be written and evaluated explicitly, with $p^T$ the transpose of $p$, as follows:

$$T = \frac{1}{2} p^T \varepsilon p = \frac{1}{2} \varepsilon_{aa} p_a^2 + \varepsilon_{ab} p_a p_b + \frac{1}{2} \varepsilon_{bb} p_b^2 = \frac{1}{2} \varepsilon_{\alpha\alpha} p_\alpha^2 + \frac{1}{2} \varepsilon_{\beta\beta} p_\beta^2.$$

The first equality gives the general expression in the matrix representation for the surplus and the remaining equalities evaluate this for the two-account example. A comparison with the expression in Table 6a shows that an additional cross term is required to account for the addition to the surplus that arises due to the substitution effect. For the eigen-baskets, this term is absent and the surplus reduces to a sum of Marshallian surpluses.

It follows that any general equilibrium problem can be reformulated in terms of a set of partial equilibrium problem. The direct surplus is non-negative, as is the kinetic energy. Therefore, the elasticity tensor must be positive-definite and, hence, an eigen-basket decomposition necessarily exists for any choice of commodities. This corroborates with the usual assumption for the matrix in economics (see, e.g., [4]).

Although in principle the tensor methods allow us address any problem in general equilibrium, this becomes when a large number of commodity types are involved. An industrialized economy, in particular, trades in an enormous number of distinct goods, making the elasticity tensor unwieldy for calculations and infeasible to assess. The same is true for mechanical systems that have a large number of degrees of freedom. In this case, other theories and methods such as thermodynamics that investigate the behavior of averages have proven to be effective in physical systems. We postpone this analogy to a follow-up paper.

## 7  Economic Engineering

In systems and control engineering, the method of analogs is exploited to model systems in a uniform manner, irrespective of their physical domain. Economic engineering extends the method to the economic domain.

Especially for linear systems, powerful techniques for analysis and control have been developed and analogies allow us to extend these to economic systems. The law of demand is a linear law. Therefore, a system consisting of demanders and linear price drivers, such as the storage and friction laws in Table 5, is a linear dynamical system. If, in addition, we assume the price elasticity and driver parameters to be constant, we obtain a class of systems called linear time-invariant, or LTI systems. These are described by linear differential equations with constant coefficients, which are widely studied in systems-and-control engineering. In this section, we apply these techniques to the description of several simple economic systems to illustrate their effectiveness.

Dynamical systems are defined by their behavior, i.e., how they respond to an external input, which is known as an exogenous variable in economics. The response or output is known as an endogenous variable. Dynamical systems are analyzed by investigating their response to standard test inputs. Impulses or steps are known as shocks in economics and the sinusoidal AC signals are referred to as seasonal or cyclical. The transient response is known as the short run and the steady state as the long run. The relationship between the input and output are given by a transfer function. In the following subsections, we describe the economic concepts of price and inventory rigidity using transfer functions. (See Figure 12.)

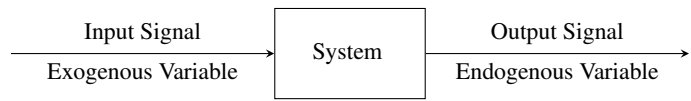

Figure 12: Block diagram specifying the effect of an exogenous input on an endogenous output.

### 7.1  First-Order Systems

In reality, there are few agents who are pure demanders, buying a constant quantity $v$ of a commodity. Instead, some degree of trade friction is always present due to handling costs. These serve to impede the flow of trade and we represent them by a damper $b$ that links the demander $m$ to the perfectly inelastic market represented by the inertial wall (see Figure 13a). Such a system is a trader, capable of both acquiring and disposing of the commodity. We subject the trader to exogenous pressure $F$ and investigate its price response. A free-body diagram for the demander gives the first-order differential equation for the trader's reservation price (see Table 13b).

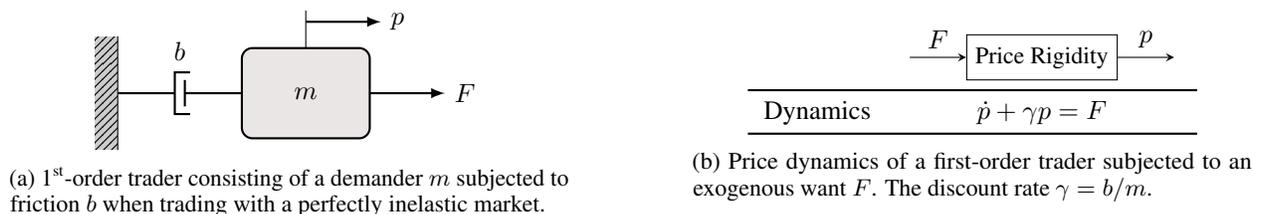

(a) $1^{\text{st}}$-order trader consisting of a demander $m$ subjected to friction $b$ when trading with a perfectly inelastic market.

(b) Price dynamics of a first-order trader subjected to an exogenous want $F$. The discount rate $\gamma = b/m$.

Figure 13: First order system illustrating price rigidity.

We first consider the free response, i.e., the price evolution of the buyer from an initial price $p_0$ onward without any exogenous price pressures. The differential equation governing the price is $\dot{p} = -\gamma p$, where the rate $\gamma = b/m$ is known as the damping rate in mechanics and the discount rate in economics (see Table 14a). In this form, the notion of $\gamma$ as a discount rate is made explicit by formulating as the percentage decline the price suffers over time. The solution $p = p_0 e^{-\gamma t}$ is known in economics as the exponential discount function.

In economics, the factor $\text{DF} = e^{-\gamma t}$ is known as the discount factor. It quantifies the agent's time preference by measuring the degree to which an agent prefers receiving the goods sooner rather than later. These are used in models for intertemporal choice that complement the usual models of choice among the available types of goods. The exponential form of the discount factor implies that the price discount depends only on the interval of time and economists consider it therefore, a time-consistent choice function. Economists typically determine the value of the discount rate $\gamma$ from data. By contrast, in economic engineering it is fully determined by the model parameters (see Table 14a).

| | Engineering | Economics |
|---|---|---|
| $\gamma = \dfrac{b}{m}$ | Damping Rate | Discount Rate |
| $\mathrm{DF} = e^{-\gamma t}$ | Decay Factor | Discount Factor |

| Price Pressure | Price Response |
|---|---|
| Free | $p = p_0 \, \mathrm{DF}(t)$ |
| Step | $p = \frac{F}{\gamma}\left(1 - \mathrm{DF}\,(t)\right)$ |

(a) System parameters and their economic analogs (above) and its time response (below).

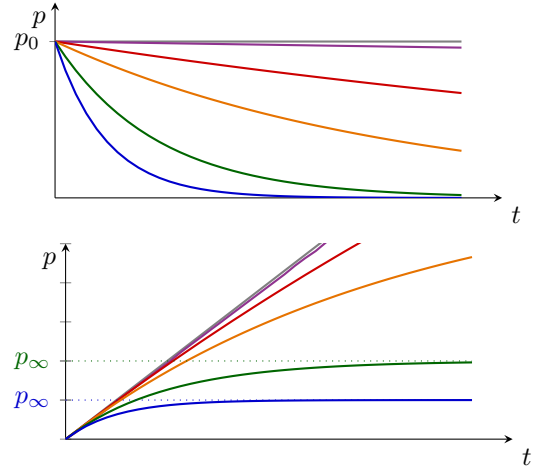

(b) Free response corresponding to an exponential discount from the initial price $p_0$ to the equilibrium price (top) and forced response to a step price-pressure shock $F$ from the equilibrium price (bottom) for discount rates $\gamma = 0$, $0.02$, $0.2$, $0.6$, $2$, and $2$.

Figure 14: Time response of a first-order agent.

Next, we consider the response of the trader to a step shock of size $F$ in price pressure. If there were no friction, the price would increase indefinitely as the agent continues to bid up the commodity to satisfy its insatiable needs. In practice, however, the handling costs serve to proportionally check the agent's wants. Ultimately, the price converges to a new equilibrium price $F/\gamma$ when the force of friction $\gamma p$ precisely balances the exogenously imposed $F$. In the short run, the price is obtained by multiplying the price pressure by a factor $\frac{1}{\gamma}(1 - \mathrm{DF})$ that econometricians call the speed-of-adjustment (see Table 14a). These effects are illustrated in Figure 14, where we graph the traders price response for various values of the discount rate.

The transfer function that determines the price action can be used as a model for price stickiness or nominal rigidity in economics. Keynes postulated that prices are resistant to change under economic shocks and do not change immediately, contrary to the neoclassical economists who argue that prices should adjust instantaneously. The economic-engineering analysis demonstrates that price stickiness is the expected behavior from a trader because it is balancing its force of demand with that of the trade friction. Because the friction force $\gamma p = bv$ increases with the level $v$ of the demand, the agent delays some of its acquisitions to avoid the higher frictional costs and finds it rational not to adjust instantaneously to the new equilibrium price.

## 7.2 Second-Order Systems

In addition to exercising demand for the commodity, second-order traders maintain an inventory. Such traders display both price and inventory dynamics simultaneously. In the following two subsections, we analyze two cases: a trader who bears transaction costs while trading, and one who bears custody costs on its inventory stock.

### 7.2.1 Inventory Dynamics

If an agent is capable of storing the commodity in addition to brokering its trades with the market, we obtain a trader whose mechanical analog is depicted in Figure 15a. We expose the agent to exogenous price pressure $F$ and investigate its inventory response. The transfer function represents the inventory rigidity of the agent. The differential equation specifying this is given in Table 15b. The inventory rigidity is now strongly dependent on the values of the parameters, and a second-order trader can act either as a broker-dealer or a buy-and-holder.

We first consider the long-run. At equilibrium, the force of demand and that of friction must vanish and, hence, any exogenous wants must be met by the convenience of the stock level. This implies that in the long run $q_\infty = F/k$. To determine the short-run transients, we notice that the dependence of the response on the discount factor form DF is identical to that of a first-order trader. However, the discount factor itself of a second-order trader is quite distinct (see Table 16c). It depends on two parameters: the natural frequency of the trade or inventory cycle $\omega_n = \sqrt{k/m}$ and, more critically, on a parameter $\zeta = \gamma/2\omega_n$ we refer to as the discount propensity (see Table 16a).

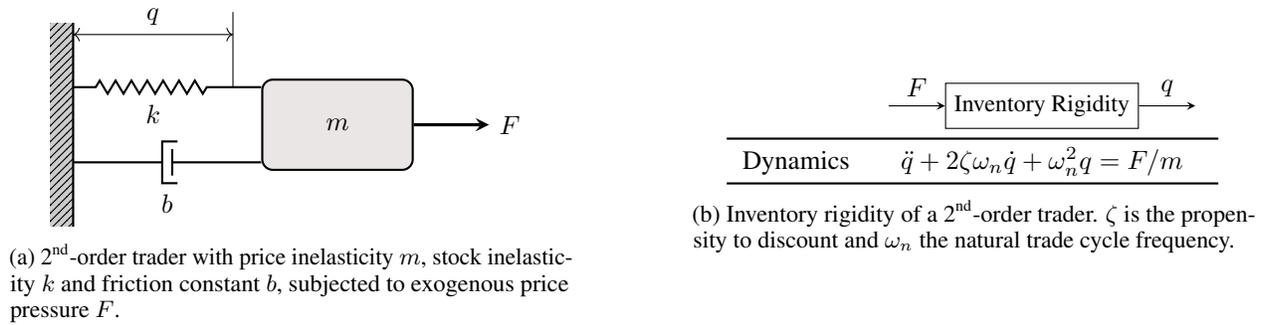

(a) 2$^{\text{nd}}$-order trader with price inelasticity $m$, stock inelasticity $k$ and friction constant $b$, subjected to exogenous price pressure $F$.

(b) Inventory rigidity of a 2$^{\text{nd}}$-order trader. $\zeta$ is the propensity to discount and $\omega_n$ the natural trade cycle frequency.

Figure 15: Second-order system.

## Cyclical Discounting

When the propensity to discount $\zeta = 0$, there is no trade friction and an equilibrium state may never be reached. The discount factor becomes a pure sinusoid with frequency $\omega_n$, implying that the trader alternates between discounting and placing a premium on the commodity. Such a trader is known as a dealer or market maker, alternately overstocking and understocking the commodity. When trade friction is introduced while keeping $\zeta < 1$, the frequency component of the discount factor is lowered to $\omega_d$ and modulated with an exponential discount at half the discount rate. Such a trader is known as a broker-dealer. This agent modulates the swings in its inventory stock to approach the equilibrium stock level in the long run.

We see that traders add storage for the same reason that a heavy body such as a car is suspended by shock absorbers, which contain springs: The storage absorbs any exogenous price shocks by selling from inventory and the springs absorb any momentum shocks from the road. The dealer controls the response characteristics by adding trade friction and in a car suspension this is done by adding damping to the shock absorbers. In control engineering, a $\zeta = 0.9$ is typical considered the best choice to obtain a rapid inventory adjustment, while keeping the maximum overshoot and rise time within acceptable levels (the orange curve in Figure 16b). Control theory, in particular transient analysis, offers various methods for tuning system parameters based on the transient analysis —a subdiscipline of control theory (see, e.g., [15])— offers methods for tuning parameters of a system to other desiderata, including rise time, maximum overshoot, settling time, etc.

## Critical Discounting

When the propensity $\zeta = 1$, the the trader ceases to act as a dealer and its behavior is the closest to that of an exponential discounter. It no longer over- and under-stocks but, rather, approaches the equilibrium stock level in an exponential manner at half the discount rate. The reason it does not do so at the full rate is because the price also moves at half the discount rate, so that the surplus moves at the sum of the two, which amounts to full rate $\gamma$. In the very short term, however, the response it dominated by a linear term $\frac{\gamma}{2}t$. The reason for this is that initially, when $v$ is still small, the trader's transaction costs are not substantial enough yet to induce the trader to deviate from the behavior of an ideal dealer.

Critical discounting is appropriate for storage facilities, such as a car gas tank, which should be filled to capacity as rapidly as possible, but which do not tolerate overstocking. Figure 16b shows that, among the curves having no overshoot, the curve for $\zeta = 1$ (green in the figure) does approach the equilibrium stock level the fastest. A prototypical control-engineering example would be door closers; it is desirable to have no overshoot at all to avoid the door slamming against the post, while simultaneously closing the door as rapidly as possible.

## Hyperbolic Discounting

When the propensity to discount moves beyond unity and $\zeta > 1$, the agent behaves as what is known as a hyperbolic discounter. This leads to correction terms that appear as hyperbolic functions. The graph of the time response in Figure 16b suggests that the agent has a higher discount rate in the very near future and lower discount rate in the more distant future.[13]

The rational for hyperbolic discounting, from the trader's perspective, is that when trading commences and the trading volume is still low, it can take advantage of the corresponding low handling costs to stock up rapidly in the short run and then slow down when these costs become significant compared to the value of the items. Although traditionally

---

[13]If we express the hyperbolic functions in terms of their definitions as a sum and difference of exponential functions, the discount factor can be rewritten as a sum of a fast and a slow exponential, explicitly showing the use of two rates of discount.

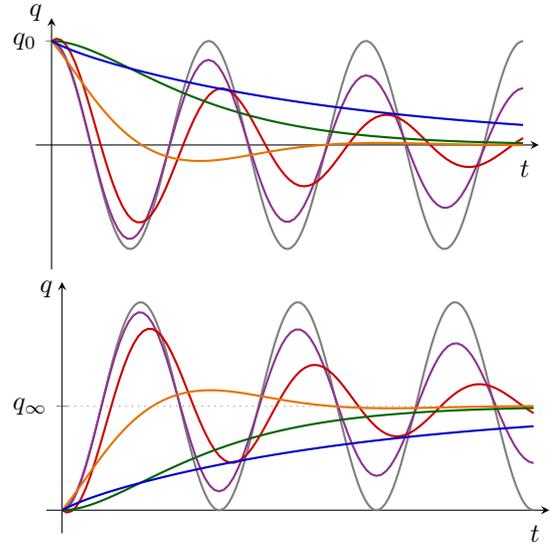

|  | Engineering | Economics |
|---|---|---|
| $\gamma = \dfrac{b}{m}$ | Damping Rate | Discount Rate |
| $\omega_n = \sqrt{\dfrac{k}{m}}$ | Natural Frequency | Trade Cycle Frequency |
| $\zeta = \dfrac{\gamma}{2\omega_n}$ | Damping Ratio | Discount Propensity |

| Price Pressure | Inventory Stock Response |
|---|---|
| Free | $q = q_0\, \mathrm{DF}(t)$ |
| Step | $q = \dfrac{F}{k}(1 - \mathrm{DF}(t))$ |

(a) System parameters and their economic analogs (above) and inventory stock response (below).

(b) Free inventory stock response (top) and forced response to price pressure (bottom) for propensities $\zeta = 0$, $0.1$, $0.3$, $0.9$, $1$, and $2$. Broker-dealers cyclically overshoot the equilibrium stock level, whereas buy-and-holders approach it in the long run.

| | Damping | Discounting | Discount Factor DF | | Frequency |
|---|---|---|---|---|---|
| $0 \leq \zeta < 1$ | Under | Cyclical | | $\cos\omega_d t + \frac{\gamma}{2\omega_d}\sin\omega_d t$ | $\omega_d = \omega_n\sqrt{1-\zeta^2}$ |
| $\zeta = 1$ | Critical | Consistent | $e^{-\gamma t/2} \times$ | $1 + \frac{\gamma}{2}t$ | $\omega_d = \omega_h = 0$ |
| $\zeta > 1$ | Over | Hyperbolic | | $\cosh\omega_h t + \frac{\gamma}{2\omega_h}\sinh\omega_h t$ | $\omega_h = \omega_n\sqrt{\zeta^2-1}$ |

(c) The discount factor, its cyclical and hyperbolic regimes, and their dependence on the discount propensity.

Figure 16: Dynamics of a second-order system.

dismissing hyperbolic discounting as irrational, economists have more recently argued that our hunter-gatherer ancestors were incentivized to consume relatively large amounts of food immediately upon finding it in order to mitigate the risks of losing it when consumption is postponed. This behavior is analogous to that of automatic vehicle braking systems, which initially brake relatively strongly, to then lighten up and come to a slow cruising halt, thus reducing the risk of sudden last-minute movements and shocks.

Hyperbolic discounting is appropriate to dampen out large shocks and to avoid the risk of overstocking due to unforeseen circumstance, while maintaining the ability to rapidly respond to the regular in- and outflow of orders. This is analogous to a typical car suspension system, where shock absorbers are actually designed to be overdamped so that the driver maintains a feel for the road from small rapid shocks, but that large shocks are dampened out.

### 7.2.2 Cost of Carry Model

Traders on the futures market profit by arbitraging the spot price of a commodity against a forward price. This can be done by storing (or shorting) the commodity at spot the price in order to dispose of (or acquire) it at some later date. When storing the commodity, traders bear what is called the cost of carry, i.e., the warehousing, insurance, and other costs involved with holding the commodity. Such a trader can be represented by the mechanical system shown in Figure 17a. By placing the damper in series with the spring rather than in parallel to it, the damper represents friction due to custody rather than to handling.

We subject the trader to an exogenous change in the desirability of the commodities and investigate its price response. The engineering diagram in Figure 17a shows that part of the total quantity acquired is lost to the custodian and does

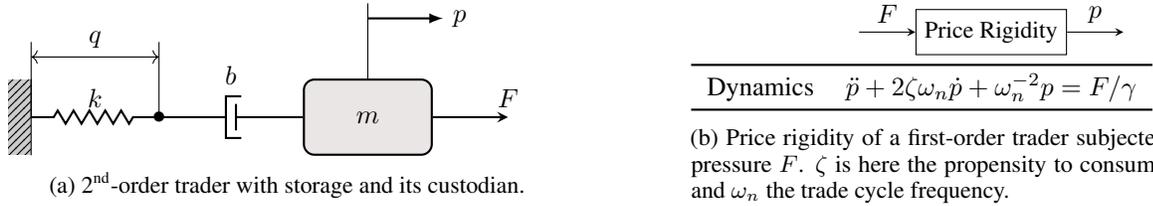

(a) 2$^{\text{nd}}$-order trader with storage and its custodian.

(b) Price rigidity of a first-order trader subjected to price pressure $F$. $\zeta$ is here the propensity to consume or carry and $\omega_n$ the trade cycle frequency.

Figure 17: Futures trader as a second-order system.

not appear in the storage. The convenience and friction forces are equal. In futures trading, this condition is referred to as "full carry" (see also [16]). The differential equation describing price rigidity is given in Table 17b.

The analysis given in Figure 15 for the second-order spot trader may be applied to the futures trader with the following modifications: the discount rate should be replaced by what is known as the carry rate and $\gamma = k/b$ is set in Table 16a. We refer to the corresponding $\zeta$ as the propensity to carry. In economics, this is also called the propensity to consume. The factor DF is known as the carry and the response is known as the cost-of-carry model. The graphs are known as futures term-structure graphs.

For the damping regimes, we distinguish between speculators for whom $\zeta < 1$, and hedgers for whom $\zeta > 1$. Traders on futures markets describe market behavior consistent with the results in Figure 16b.

When the market is dominated by hedgers, one speaks of a normal market. Hedgers are willing to bear the cost of storage in order to take later delivery. As a consequence, the price advances consistently towards the deferred contracts consistent with the price responses in Figure 16b for any $\zeta \geq 1$. This is known as normal backwardation.

When speculators take over the market, however, the price overshoots. When it moves back, the market is said to be inverted. The explanation is that, when inventories are low and a shortage is felt, those speculators in need bid up nearby contracts to a momentary premium over deferred ones. The analogous mechanical statement is that "a spring (inventory) in compression (shortage) generates the acceleration (need) that forces (bids up) the momentum (price) to overshoot (a premium)." This is borne out by the price responses in Figure 16b for carry propensities $\zeta < 1$. In theory, speculative markets switch cyclically between being in normal backwardation and inverted at the frequency $\omega_d$. Because persistent fluctuations are rarely seen, we can conclude that, in practice, inverted markets have propensities close to $\zeta = 0.9$ of the orange graph, as it settles rapidly enough to dampen out all but the the first of the inversions.

## 7.3 Dynamical Systems

In this section, we briefly overview several ways the analysis for the first and second-order systems of the previous subsections can be generalized.

### General Second-Order Systems

The spot and futures traders can be consolidated into one general second-order trader who both incurs handling and carrying costs. Only in the very short time, i.e. within a cycle, is the behavior influenced by relative strength of the handling costs with respect to the carrying cost as the agent attempts to optimally balance these. In the medium or long term, i.e. over more than at least several cycles, the behavior of such a trader does not deviate substantially from that of a spot or futures trader.

### Higher-Order Linear Systems

In general, an economic system may display several natural trade cycles. For instance, economists identify at least four different cycles in the economy. The period of a cycle can range from four years for the inventory cycle to an approximately 50 years for the technology cycle.

For instance, economists identify at least four different cycles in the economy, whose periods ranging from the 4-year period of the inventory cycle, to an approximately 50-year period of the technology cycle. This means that the economy is at least an eight-order system, consisting of four interacting second-order systems.

### Nonlinear Systems

As in mechanics, nonlinear force laws occur more often than not in economics. It is easy to imagine a storage facility overflowing and it is equally imaginable that a spring will break when stretched enough. Although general solution methods are lacking, sophisticated methods have been developed for nonlinear system analysis in control engineering, and these are equally applicable to economic systems.

Economic texts typically picture demand and supply curves as convex curves, suggesting the presence of a nonlinear demander. However, an analogous nonlinear inertial element cannot be entertained within the confines of Newtonian mechanics, since mass is a constant, independent of velocity. Therefore, if price elasticity appears to be change, its dependence on quantity demanded is implicit via time. A well-known example in mechanics is that of a rocket ship whose mass decreases over time while it picks up speed due to the emission of exhaust fumes. An analogous economic situation would involve a gradual rotation of the demand curve as quantity demanded increases. When measuring price elasticity, it is critical that the demander is sufficiently isolated from any exogenous effects that may affect the reading.

### 7.4 Control Theory

An important advantage of dynamical-systems models is that their behavior can be tuned using the methods and tools of control theory. In economic applications, we think of the controller as a manager —such as a policy maker or a financial regulator— of the economic system. Its policy and executive decisions constitute the controller actions. Its objectives or desired price (or stock) level represent the controller setpoint. In Figure 18 we show how feedback control can be used to design a price (or inventory) management system. The observed price $p$ (or stock $q$) is compared to its objective $\bar{p}$ (or $\bar{q}$), producing the error signal $e$, which quantifies the deviation from the objective. Based on that, management intervenes by exercising corrective price actions $F$. The price (or stock) rigidity serves as the transfer function relating these actions to adjustments in the actual level, which is then fed back for further corrective actions.

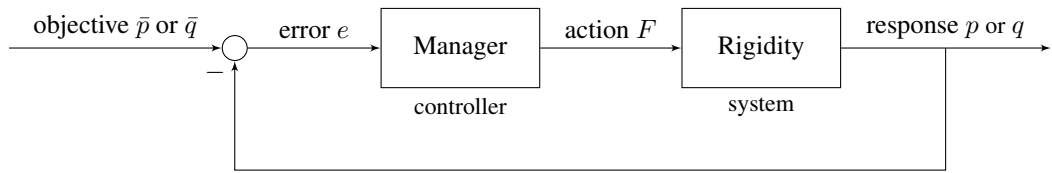

Figure 18: Feedback control for price or inventory management.

The design of a suitable control law depends on the specific policy targets that are in place. A Proportional-Integral-Derivative (PID) controller, for instance, can be tuned to meet response time targets while eliminating any steady-state errors. Originally modelled after the behavior of helmsmen on ships, PID control actions are both effective and intuitive and they are widely used in industry. We expect these features to carry over to economic applications. Many other control schemes exist for dynamical systems, each with there unique features which may be relevant for managers, financial regulators, or policy makers.

## 8  Conclusions

This paper is one in a series of publications that will provide a theoretical foundation for economic engineering. Our purpose herein is to develop a theory based on an economic analog to Newtonian mechanics. By way of conclusion, we evaluate our development by comparing it to existing treatments in both economic and engineering literature. Finally, we identify several limitations of the theory and indicate how we intend to address these in our forthcoming publications of the series.

The crucial element in an engineering system is an *inertial* element, i.e., a mass in mechanical systems (or an inductor in electrical systems). In the paper, we recognize *demand* in economics as analogous to inertia. we consider this our critical insight, from which all else follows. It complements existing approaches for modelling economic systems in engineering literature, where inertial elements are systematically missing.[14]

In engineering, the force concept is central to conceptualization and modelling efforts. Building on our concept of demand, we define the *force of demand* as analogous to Newton's definition of the *force of inertia*. Any economic force is then calibrated by comparing it to this force of demand, analogous to Newton's development of a mechanical force. In the following two paragraphs, we compare the relative advantages of the use of force in economic modelling over econometric methods and the theory of demand.

Econometric models rely on *correlations* uncovered using *statistical* theories. In contrast, forces lead to *causal* models based on economic *laws*. [15] Such models have several important advantages. First, they are more reliably predictive since correlations may break down. Second, the models are more readily interpretable since the laws give economic meaning to any parameters. Third, less data is required to identify the model since only a limited set of parameters needs to be determined to generate an entire time series of price and stock adjustments. These advantages are especially important for modelling and the design of highly complex systems.

---

[14]To wit, the hydraulic diagrams in Forrester's [6] system dynamics lack an inertial paddle wheel (see Figure 20).

[15]In the paper, we show how, analogous to Newton's second law of motion, the law of demand identifies the economic force as the cause for a change in behavior.

In both classical and neoclassical theories of demand, prices are assumed to adjust nearly instantaneously after a shock. The role of force is thus restricted to one that is *static*. This contrasts with the *dynamic* nature of the Newtonian economic force, which details how prices and stocks evolve over time. Although Keynesians do distinguish between a short and a long run, a dynamic economic force specifies behavior over any time span, no matter how short it is and no matter when it occurs. Dynamic forces are particularly valuable when modeling volatile economic conditions, where predictions of short-term transient price movements are of the essence. In addition, it allows us to model systems that never attain equilibrium, such as those with recurring business cycles or persistent economic growth paths.

Although the Newtonian theory is particularly useful in economic engineering practice, it has two important limitations; one theoretical and one practical. Our forthcoming publications in the foundational series address these by applying analytical mechanics —i.e., Lagrangian and Hamiltonian mechanics— to economics. Analytical mechanics provide a definitive set of theoretical foundations for economic engineering: Lagrangian mechanics from the perspective of the individual in terms of utility maximization and Hamiltonian mechanics from the perspective of a business in terms of flow of surplus. For engineering practice, they provide a unified framework wherein we can extend economic engineering beyond trade in commodities to areas such as production, economic growth, uncertainty, etc. With the completed series, we thus aim to establish comprehensive theoretical foundations for economic engineering on a par with those for mechanical engineering.

# Appendices

## A   Mobility vs Impedance Analogies

In this appendix, we overview the two analogy variants and their connection with the choice in economics among flow and stock variables on the one hand, or various marginal quantities on the other hand. We also include the electrodynamic analogy and a hydrodynamic analogy that is consistent with thinking in system dynamics.

Newton's development of mechanics is consistent with the mobility analogy. For economics, this implies that values flow (in or out) and its conjugate demand extends (or contracts). Newton introduced the overdot notation to write this flow as $\dot{p}$.[16] In Figure 19, we show how the free-body diagram creates the picture of a flow of value and, consequently, how the physical demand can be seen to be a marginal change in level rather than a "quantity demanded" (see Table 19a).

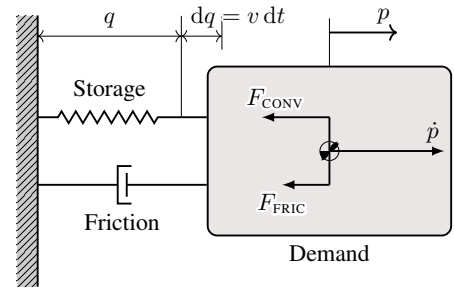

|  |  | **Mechanics** | **Economics** |  |
|---|---|---|---|---|
| *Kinematics* | $q$ | Position | Balance | $+$ |
|  | $v$ | Velocity | Demand Level | \| ↑ |
| *Dynamics* | $p$ | Momentum | Value | Dr |
|  | $F$ | Force | Want | → |

(a) Wants as a vector, representing a flow of value, and demand as a covector, representing an marginal rise in the demand level. Values are debited or credited (Dr or Cr) and the raising of the balances is positive or negative (+ or -).

(b) The free-body diagram pictures the wants as vectors representing flow of value from the demander to the forces of convenience and friction. The demand level is pictured as the component of the covector $dq = v\,dt$.

Figure 19: The Mobility analogy

The theory of demand is mostly consistent with the impedance analogy. Properly $v$ is a *flow* quantity demanded, in units of the commodity per period of time and, similarly, the term "volume" as used for securities is properly the volumetric flow rate. The analogy is consistent with electrodynamics and hydrodynamics (see Figure 20). In an electrical circuit, the flow of electrical current results in a voltage drops and, in a hydrodynamic circuit, the fluid flow results in a pressure drop. Both the circuits in Figure 20 can be read off economically: the commodities flow from (or to) the demander into or (out of) storage, and then via the handler to the market. Each element results in a rise (or drop) in desirability; specifically, the marginal price rise (or drop) $dp$ by the demander is determined by the inducement $kq\,dt$ from the convenience of storing the commodities net of inducements $bv\,dt$ provided to the handler.

In engineering, the flow and marginal quantities of economics are known as through and across variables (see Table 9). For instance, forces go through an element, while pressures or voltages are measured across and this imagery can be applied, mutatis mutandis, to economic forces and desirabilities. From a differential-geometric perspective, a flow quantity is represented by a vector and a marginal quantity by a covector or 1-form. The vector gives the both the magnitude and the direction of the flow in the usual manner. The covector does this for the marginal quantities in terms of a pair of level lines with a flag for the up direction. Table 9 shows how these naturally act on each other through the interior product to yield the instantaneous power $Fv$, i.e., the allocation rate of economic surplus.

The energy variables are found by integrating the power variables. For economics, the surplus is found by integrating- or accruing the allocations. Due to the conjugacy of the power variables, the integral of the surplus flow becomes a regular line integral in both cases (see Table 9). The economic interpretation is, however, quite distinct.[17] On the one hand, an inflow (or outflow) adds up to an accumulation (or decumulation or reduction). Accountants refer to this as a debit (or credit) when it concerns an asset account and use the Dr/Cr symbolism in Table 9 to indicate the net direction. In Section 3.3, this terminology was used to paraphrase the third law in terms of an equality of total credit and debit amounts. Economists refer generically to an integral of a flow as a stock variable, but this terminology is unnatural and

---

[16]Suggestively, Newton referred to the notation as a fluxion.

[17]Historically, the notation was also distinct. Newton developed the box notation for the integral of a flow, as in his inherent force $p = \Box\dot{p}$, suggesting that the force fills up the mass as it were while it flows in it. Leibniz's notation, as in $p = \int \dot{p}\,dt$, which is used exclusively nowadays, is consistent with the marginal interpretation.

|          |     | **Electrodynamics** | **Hydrodynamics**    | **Economics**     |     |
|----------|-----|---------------------|----------------------|-------------------|-----|
| *Kinematics* | $q$ | Charge          | Volume               | Acquisitions      | Dr  |
|          | $v$ | Current             | Volumetric Flow Rate | Quantity Demanded | →   |
| *Dynamics*   | $p$ | Flux Linkage    | Momentum             | Price             | +   |
|          | $F$ | Voltage             | Pressure             | Desirability      | ⇂↾  |

(a) Commodity flows as a vector and the desirability as a covector, representing an infinitesimal rise of the level of demand. The acquisitions are debited (or credited) and the raising of the price is positive (or negative).

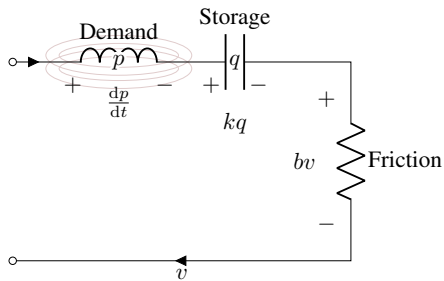

(b) Electrical circuit. The electric current represents the flow of goods and goes through the elements. The voltage drop represents their desirability and the price movement constitutes the sum of the convenience and friction forces.

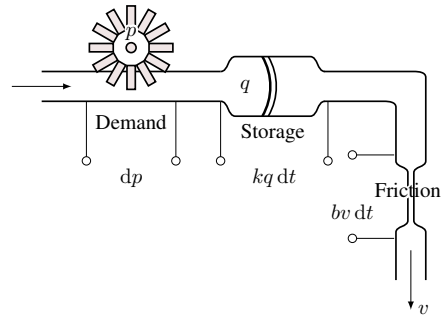

(c) Hydraulic circuit. The fluid flow represents the flow of goods and passes through the elements. The price drops and the inducements that cause these are given explicitly as covectors, analogous to the representation of the pressure drops.

Figure 20: The impedance analogy

| **Differential Geometry** | | **Engineering** | | **Economics** | |
|---|---|---|---|---|---|
| → | Vector | Power | Through | Surplus Flow | Flow |
| ⇂↾ | Covector | | Across | | Marginal |
| Dr/Cr | Number | Energy | Accumulation | Surplus | Stock |
| +/- | | | Extension | | Level |

Table 9: Categorization of variables in engineering and economics. Covectors are represented as a pair of straightened-out level lines with a flag for the up direction. The combination Dr/Cr stands for credit or debit and derives from accounting, with debit representing an inflow and credit an outflow.

confusing since such an integral is not necessarily equal to a stock amount that is registered in a state variable. On the other hand, marginal increases (or decreases) add up to an extension (or contraction). Typically, one associates the upward (downward) direction with the $+$ $(-)$ sign. In this manner, it is natural to speak of a high or elevated price in the impedance analogy and similarly for the level of demand in the mobility analogy.

In the impedance analogy, multiple accounts requiring higher dimensional commodity spaces are dealt with by adding a closed circuit for each additional dimension. Although these circuits cannot exchange commodities —i.e., electrical charge or hydraulic fluid— they can influence each other by communicating prices —i.e., transferring magnetic flux or momentum. In electrodynamics, this is known as mutual inductance and leads to tensors that can be decomposed into orthogonal directions in the same manner as presented in Section 6.4.